\documentclass[11pt, oneside, a4paper]{article}
\usepackage{mcite}
\usepackage{cite}
\usepackage{latexsym}
\usepackage{amssymb}
\usepackage{amsmath}
\usepackage{amsthm}
\usepackage[applemac]{inputenc}
\newcommand{\be}{\begin{eqnarray}}
            \newcommand{\ee}{\end{eqnarray}}
           \newcommand{\eel}[1]{\label{#1}\end{eqnarray}}
\newcommand{\e}[1]{\label{eq:#1}\end{eqnarray}}
     \newcommand{\eg}{{\em e.g.\ }}
            \newcommand{\ie}{{\em i.e.\ }}
            \newcommand{\ga}{{\gamma}}
 
            \newcommand{\la}{{\lambda}}

\newcommand{\del}{{\delta}}

\newcommand{\bphi}{\bar{\phi}}
\newcommand{\brho}{\bar{\rho}}
\newcommand{\bbeta}{\bar{\beta}}
   
\newcommand{\dx}{{\dot{x}}}

 \newcommand{\cP}{{\cal{P}}}

\newcommand{\cW}{{\cal{W}}}
\newcommand{\cB}{{\cal{B}}}

\newcommand{\cQ}{{{\cal Q}}}
\newcommand{\caR}{{{\cal R}}}
\newcommand{\cp}{{{ p\!\!\!\slash}}}
\newcommand{\cdif}{{{ \partial\!\!\!\slash}}}

           \newcommand{\ra}{{\rightarrow}}

  \newcommand{\tI}{\tilde{{I}}}
\newcommand{\tphi}{\tilde{{\phi}}}

\newcommand{\trho}{\tilde{{\rho}}}
\newcommand{\tve}{\tilde{{\varepsilon}}}
\newcommand{\tPi}{\tilde{{\Pi}}}
\newcommand{\txi}{\tilde{{\xi}}}
  \newcommand{\tF}{\tilde{F}}
   \newcommand{\tPsi}{\tilde{\Psi}}

\newcommand{\hps}{\hat{\psi}}

\newcommand{\dpsi}{{\dot \psi}}

\newcommand{\cC}{{\cal C}}

            \newcommand{\beq}{\begin{quote}}
            \newcommand{\eq}{\end{quote}}
            
            \newcommand{\al}{\alpha}
            \newcommand{\ben}{\begin{enumerate}}
            \newcommand{\een}{\end{enumerate}}
            \newcommand{\bit}{\begin{itemize}}
            \newcommand{\ei}{\end{itemize}}
        \newcommand{\nn}{\nonumber}
            \newcommand{\rl}[1]{(\ref{eq:#1})}
            \newcommand{\edfl}[1]{\Label{#1}\end{df}}

\newcommand{\vb}{{\cal h}}
\newcommand{\hb}{{\cal i}}

\newcommand{\cG}{{\cal G}}

\newcommand{\cA}{{\cal A}}
\newcommand{\cF}{{\cal F}}
\newcommand{\cE}{{\cal E}}
\newcommand{\cX}{{\cal X}}
\newcommand{\cR}{{\cal R}}
\newcommand{\cro}{{\cal R}}

\newcommand{\dif}{{\partial}}
\newcommand{\half}{\frac{1}{2}}

\begin{document}
\begin{titlepage}
\begin{center}
{\LARGE\bf Lagrangian higher spin field theories from the O(N) extended supersymmetric particle}\end{center}
\vspace*{3 mm}
\begin{center}
\vspace*{3 mm}

\begin{center}
Robert Marnelius
 \\ \vspace*{7 mm} {\sl
Department of Fundamental Physics\\ Chalmers University of Technology\\
S-412 96  G\"{o}teborg, Sweden}\end{center}
\vspace*{10 mm}
\begin{abstract}
The wave function in the quantum theory of the O(N) extended supersymmetric particle model describes a massless free field with spin N/2. This quantum theory is here exactly solved in terms of gauge fields in arbitrary even dimensions using only the basic quantum operators which include graded external differentials, trace operators, index structure operators and their duals. The resulting equations for the gauge fields are of first (N odd) or second order (N even) and are shown to be generalized (Fang)-Fronsdal equations  which are fully gauge invariant since they include compensator fields in a natural way. Local gauge invariant actions are first  derived in analogy with the derivation by  Francia and Sagnotti in the symmetric case. Then a minimal formulation is given within which it is easy to set up gauge invariant actions and here appropriate actions for the above equations are proposed. \\ \indent In a second part it is shown that there exist projection operators from the states of the field strengths (wave functions) to Weyl states (Weyl tensors) expressed only in terms of the trace operators and their duals for integer spins or in terms of the gamma trace operators and their duals for half-integer spins. These Weyl states are not just fully gauge invariant but also invariant under generalized Weyl transformations. If one lets the equations be determined by the actions defined to be the squares of these Weyl tensors then the theory is conformally invariant and the equations are of order N for the gauge fields  defined as in the exact theory. In d=4 this higher order conformal theory fits into the framework for conformal higher spin theories set up by Fradkin, Tseytlin and Linetski.
\end{abstract}\end{center}\end{titlepage}

\tableofcontents

\setcounter{equation}{0}
\section{Introduction}
In this paper I show that the  properties of free field theories for massless particles of arbitrary spins may be derived from the $O(N)$-extended supersymmetric particle model \cite{Gershun:1979fb,Howe:1988ft,Marnelius:1988ab}. This includes the first and second order (Fang)-Fronsdal equations \cite{Fronsdal:1978rb,Fang:1978wz} in antisymmetric form as well as the later discovered fully gauge invariant first and second order equations also in antisymmetric form. Fully gauge invariant Lagrangians are then first derived in analogy with the ones by  Francia and Sagnotti \cite{Francia:2005bu} in the symmetric case. Then I give a minimal formulation within which it is easy to give manifestly gauge invariant Lagrangians and here the forms of the appropriate actions are given. In a second part I then further develop the  free conformal higher order field theories presented in \cite{Marnelius:2008er} which
are extracted from the $O(N)$-extended supersymmetric particle model. Here the actions  determine the equations which then are both fully gauge invariant and Weyl invariant under generalized Weyl transformations. In $d=4$ these conformal theories  fit into  the general framework set up by Fradkin, Tseytlin and Linetski \cite{Fradkin:1985am,Fradkin:1989md,Fradkin:1990ps} apart from the use of a different but equivalent representation.

In part {\bf I} I treat the exact quantum theory of the $O(N)$ extended supersymmetric particle model. That the quantized wave function of the $O(N)$-extended supersymmetric particle model satisfies field equations for particles with spin $s=N/2$ (in $d=4$) were independently proposed in \cite{Gershun:1979fb,Howe:1988ft,Marnelius:1988ab}. (A representation independent proof is given in \cite{Marnelius:1990de}.) The treatment here is a direct continuation of the treatment of the $O(N)$-particle given in \cite{Marnelius:1988ab} except that I here will make use of an equivalent Dirac quantization. Furthermore, I generalize the $d=4$ treatment in \cite{Marnelius:1988ab} to arbitrary even dimensions $d$. It turns out that the quantum treatment of \cite{Marnelius:1988ab} provides for a powerful compact notation which essentially is a multilinear form language applied to flat Minkowski space. In addition  it contains also duality transformations in a simple fashion. (This language may be compared to the one in  \cite{DuboisViolette:1999rd,DuboisViolette:2001jk,Henneaux:2008ee,Bekaert:2002dt,Bekaert:2003az,Bekaert:2006ix}.) Anyway the main advantage of this formulation is that it allows for a detailed treatment without writing down any tensor equations. The equations for the wave functions are here easy to solve in terms of equations for gauge fields, where the latter are first or second order equations which are fully gauge  invariant by means of compensator fields which appear naturally.  (Compensator fields  were directly derived in \cite{Francia:2002aa,Francia:2002pt,Francia:2005bu} but  previously  they appeared as \eg something extracted from string theory \cite{Bengtsson:1986ys,Pashnev:1998ti,Sagnotti:2003qa}.) The initial part of the derivation here is  inspired by the one performed in \cite{Bandos:2005mb} although the actual formulas look different (cf also \cite{Labastida:1986ft,Labastida:1986zb,Bekaert:2002dt,Bekaert:2006ix} and the papers above). Notice that only in $d=4$  do I have purely symmetric gauge fields. The corresponding equations for the dual gauge fields are obtained in a simple fashion.  Fully gauge invariant actions in terms of gauge fields, general compensator fields and Lagrange multipliers are derived following the derivation of the minimal actions in \cite{Francia:2005bu} within the symmetric formulation and for symmetric gauge fields.  However, it is doubtful that they without modifications yield  the correct equations. I define therefore  a minimal formulation  where the compensator fields have their most simple form allowed by complete gauge invariance. Here it is easy to set up fully gauge invariant actions. Within a natural class of actions I then propose the appropriate ones for the above equations in their minimal form.

In part {\bf II} I treat a higher order conformal field theory which is extracted from the exact quantum theory in part I by retaining the original gauge invariant wave functions (field strengths) and their tensor properties but removing all equations of motion. It was presented in \cite{Marnelius:2008er}. The key objects here are generalized Weyl tensors or Weyl states. 
These states are just projections from the original gauge invariant states and here the explicit form of these projection operators are given. Such Weyl states  are defined for  both integer and half-integer spins generalized to arbitrary even dimensions. They are expressed in terms of 
generalized traceless Weyl tensors. This is one property of the projected Weyl states. Another is that they also are
invariant under generalized Weyl transformations.   Conformally invariant actions may be defined as the scalar product of these Weyl states (the square of the Weyl tensors).  However, the equations from these actions are 
 not the same as the equations removed to start with. 
 Instead of being of first or second order
 they are of order $N=2s$ for any spin $s$ in arbitrary even dimensions. 
 That the equations are of higher order is a drawback  which also blurs the spin concept. However, the actions are fully gauge invariant and possible to quantize although it remains to find consistent unitary solutions. The main motivation for this formulation is that it definitely seems to allow for conformally invariant  interactions in a Lagrangian form. (The evidence so far is the results of the papers \cite{Fradkin:1989md,Fradkin:1990ps,Arvidsson:2006}.)
The treatment here generalizes  \cite{Marnelius:2008er} in the following aspects: The forms of the Weyl tensors (states) are given for both integer and half-integer spins in arbitrary even dimensions. Furthermore, duality properties are given and many new details.

The paper is organized as follows: In part {\bf I} section 2 I present  the quantum $O(N)$ extended supersymmetric particle model given in \cite{Marnelius:1988ab}. In section 3 I show then how some equations are naturally solved in terms of gauge states (fields). There are also dual equations which may be solved in terms of dual gauge states (fields). Then I show that the remaining equations are solved by gauge invariant second order equations for the gauge fields, or  their duals in the integer spin case. In the half-integer spin case the corresponding equations are of first order.  In section 4 I derive some properties of these equations which in essence are similar to what have been obtained in other formulations. The (Fang)-Fronsdal limit shows agreement with the expected properties in $d=4$ although the formulation is different. In section 5 I give gauge invariant actions which are of (first) second order in the gauge fields in the (half-)integer spin case and which also contains general compensator fields and Lagrange multiplier fields. (These actions have a similar structure to the actions given by Francia and Sagnotti in \cite{Francia:2005bu} for the symmetric case.) However, without modifications they seem not to yield  the correct equations here. In section 6 the minimal formulation is presented in which manifestly gauge invariant expressions  are easy to write down. Here the forms of the appropriate Lagrangians are proposed. In section 7 I then discuss  the BRST-quantization of the $O(N)$ extended supersymmetric particle model. I end part I with some remarks in section 8.  In part II I treat and considerably expand the higher order theory given in \cite{Marnelius:2008er}. In section 9 I outline the results of the following sections 10 and 11, sections in which I give a rather detailed derivation of the Weyl states for integer and half-integer spins. This includes also their duality properties. In section 12 I give the conformally invariant actions  and then I end part II by some final remarks in section 13. In section 14 I compare the two theories presented in part I and II. In  appendices A and B I derive the duality properties from the quantum theory, and in appendix C some of the Lagrangian equations in section 5  are displayed. 

\part{The exact quantum theory}

   \setcounter{equation}{0}
   \section{The Lagrangian of the $O(N)$ extended supersymmetric particle model and its quantum properties}
   The Lagrangian of the massless $O(N)$ extended supersymmetric particle model may be written as follows
   (repeated indices are summed over) \cite{Gershun:1979fb,Howe:1988ft ,Marnelius:1988ab}
   \be
   &&L(\tau)={1\over 2v}\big(\dx-i\la_j\psi_j\big)^2+\half i\psi_j\cdot\dpsi_j-\half i f_{kl}\psi_k\cdot\psi_l,
   \e{101}
   where $x$ is the spacetime coordinate, $\psi_j$, $j=1,\ldots,N,$ are real odd Grassmann variables, and $\la_j, f_{kl}(=-f_{lk}), v$ are real Lagrange multipliers where $\la_j$ is Grassmann odd and $f_{kl}, v$ are Grassmann even. (There is a Lagrangian for each integer value of $N$. However, since I describe the theory for a generic $N$ I suppress this $N$ dependence in the Lagrangian \rl{101} and the Hamiltonian below.) The Hamiltonian formulation of \rl{101} determines the quantization, and it gives rise to the following constraints ($H=c_a\chi_a$):
   \be
   &&\chi_a=0,\quad \chi_a\equiv\Big\{p^2,p\cdot\psi_j,\psi_k\cdot\psi_l\Big\},
   \e{102}
   where $p$ is the momentum. After quantization the  operators $x$, $p$, and $\hps_j$ satisfy the (anti)commutation relations (hats on $x$ and $p$ are suppressed)
   \be
   &&[x^{\mu}, p_{\nu}]_-=i\del^{\mu}_{\nu},\nn\\
   &&[\hps_j, \hps_k]_+=\eta^{\mu\nu}\del_{jk},
   \e{103}
   where $\eta$ is the Minkowski metric. A straightforward Dirac quantization yields    the equations 
      \be
   &&\hat{\chi}_a|\;\;\;\hb=0,
   \e{1031}
   where $\hat{\chi}_a$ are the corresponding operators to $\chi_a$ in \rl{102}.  (\rl{1031} are the corresponding quantum constraints to the classical ones in \rl{102}.)  Since $\sqrt{2}\hps_j$ are naturally represented as $\ga$-matrices, the wave functions from \rl{1031} are multispinors (cf the treatment in \cite{Gershun:1979fb,Howe:1988ft}). In order to get tensor fields one has to follow the procedure in \cite{Marnelius:1988ab} and define fermionic creation and annihilation operators by
   \be
   &&b_r^{\mu}\equiv{1\over\sqrt{2}}\Big(\hps_r^{\mu}+i\hps_{r+n}^{\mu}\Big),\nn\\
   &&r=1,\ldots,n\equiv[s],
   \e{104}
   where $[s]$ is the integer part of the spin $s\equiv N/2$. These $b$-operators satisfy the anticommutation relations
   \be
   &&[b_r^{\mu}, b_q^{\dag\nu}]_+=\eta^{\mu\nu}\del_{rq},\nn\\
   &&[b_r^{\mu}, b_q^{\nu}]_+=[b_r^{\dag\mu}, b_q^{\dag\nu}]_+=0.
   \e{105}
   Notice that $b$-operators only exist for $s\geq1$.
   
   For integer spins $s$ (even $N$)   the general ansatz for a quantum state has the form
 (the sum is over all possible values of the integers $n_j$, $0\leq n_j\leq d$, where $d$ is the dimension of spacetime)
\be
&&|F\hb=\nn\\
&&\sum_{n_j}F_{\mu_1\cdots \mu_{n_1};\nu_1\cdots \nu_{n_2}; \rho_1\cdots \rho_{n_3}; \cdots;     \la_1\cdots \la_{n_s}}(x)|0\hb_{(p)}^{\mu_1\cdots \mu_{n_1};\nu_1\cdots \nu_{n_2};\rho_1\cdots \rho_{n_3};\cdots;\la_1\cdots \la_{n_s}},\nn\\
\e{106}
where 
\be
&&|0\hb_{(p)}^{\mu_1\cdots \mu_{n_1};\nu_1\cdots \nu_{n_2};\rho_1\cdots \rho_{n_3};\cdots;\la_1\cdots \la_{n_s}}=\nn\\
&&\qquad\qquad|0\hb^{\mu_1\cdots \mu_{n_1};\nu_1\cdots \nu_{n_2};\rho_1\cdots \rho_{n_3};\cdots;\la_1\cdots \la_{n_s}}|0\hb_{(p)},\nn\\
&&|0\hb^{\mu_1\cdots \mu_{n_1};\nu_1\cdots \nu_{n_2};\rho_1\cdots \rho_{n_3};\cdots;\la_1\cdots \la_{n_s}}
\equiv  {1\over\sqrt{n_1!n_2!\cdots n_s!}}\times\nn\\
&& b_1^{\mu_1\dagger}\cdots b_1^{\mu_{n_1}\dagger}b_2^{\nu_1\dagger}\cdots b_2^{\nu_{n_2}\dagger}b_3^{\rho_1\dagger}\cdots b_3^{\rho_{n_3}\dagger}\cdots\cdots b_s^{\la_{1}\dagger}\cdots b_s^{\la_{n_s}\dagger} |0\hb,\nn\\
&& b_j^{\mu}|0\hb=0,\quad p_{\mu}|0\hb_{(p)}=0.
\e{107}
$F_{\ldots}(x)$ are arbitrary fields (wave functions). Reality of $F$ s not required by the quantum theory but may be imposed.
Some basic properties of  states like \rl{106} are given in appendix A. (For $s=0 (N=0)$ $F$ is a scalar. \rl{101} is the Lagrangian of a scalar particle.)

For half-integer spins $s$ (odd $N$) the general ansatz is
\be
&&|\Psi\hb=\nn\\
&&\sum_{n_j}\Psi^{\al}_{\mu_1\cdots \mu_{n_1};\nu_1\cdots \nu_{n_2}; \rho_1\cdots \rho_{n_3}; \cdots;     \la_1\cdots \la_{n_{[s]}}}(x)|0\hb_{\al}^{\mu_1\cdots \mu_{n_1};\nu_1\cdots \nu_{n_2};\rho_1\cdots \rho_{n_3};\cdots;\la_1\cdots \la_{n_{[s]}}},\nn\\
&&|0\hb_{\al\;(p)}^{\mu_1\cdots \mu_{n_1};\nu_1\cdots \nu_{n_2};\rho_1\cdots \rho_{n_3};\cdots;\la_1\cdots \la_{n_{[s]}}}=\nn\\
&&\qquad\qquad |0\hb^{\mu_1\cdots \mu_{n_1};\nu_1\cdots \nu_{n_2};\rho_1\cdots \rho_{n_3};\cdots;\la_1\cdots \la_{n_{[s]}}}|\al\hb|0\hb_{(p)},\nn\\
\e{108}
where $\al$ is a spinor index. Notice that for half-integer spins one has apart from the $b$-operators \rl{104} one remaining $\hps$-operator satisfying \rl{103}. The latter may be decomposed to yield the spinor state $|\al\hb$ satisfying the scalar product (see appendix in \cite{Marnelius:1988ab} which trivially is generalizable to arbitrary even dimensions)
\be
&&\vb\al|\beta\hb=\ga^0_{\al\beta}.
\e{1081}
$\sqrt{2}\hps$ is then represented by a $\ga$-matrix as follows
\be
&&\ga^{\mu}_{\al\beta}=\ga^0_{\al\ga}\vb\ga|\sqrt{2}\hps^{\mu}|\beta\hb.
\e{1082}

That the states $|F\hb$ and $|\Psi\hb$ in \rl{106} and \rl{108} depend on $s$ is suppressed. However, this $s$ dependence becomes explicit in the index structure of the corresponding wave functions which are
\be
&&F^{\mu_1\cdots \mu_{n_1};\nu_1\cdots \nu_{n_2}; \rho_1\cdots \rho_{n_3}; \cdots;     \la_1\cdots \la_{n_s}}(x)=\nn\\&&\qquad\qquad=\:^{\mu_1\cdots \mu_{n_1};\nu_1\cdots \nu_{n_2}; \rho_1\cdots \rho_{n_3}; \cdots;     \la_1\cdots \la_{n_s}}\vb x|F\hb,\nn\\
&&\Psi_{\al}^{\mu_1\cdots \mu_{n_1};\nu_1\cdots \nu_{n_2}; \rho_1\cdots \rho_{n_3}; \cdots;     \la_1\cdots \la_{n_{[s]}}}(x)=\nn\\&&\qquad\qquad=\:^{\mu_1\cdots \mu_{n_1};\nu_1\cdots \nu_{n_2}; \rho_1\cdots \rho_{n_3}; \cdots;     \la_1\cdots \la_{n_{[s]}}}{}_{\al}\vb x|\Psi\hb,\nn\\
\e{1083}
where
\be
&&|x\hb^{\mu_1\cdots \mu_{n_1};\nu_1\cdots \nu_{n_2}; \rho_1\cdots \rho_{n_3}; \cdots;     \la_1\cdots \la_{n_{[s]}}}=|0\hb^{\mu_1\cdots \mu_{n_1};\nu_1\cdots \nu_{n_2}; \rho_1\cdots \rho_{n_3}; \cdots;     \la_1\cdots \la_{n_{[s]}}}|x\hb,\nn\\
&&|x\hb_{\al}^{\mu_1\cdots \mu_{n_1};\nu_1\cdots \nu_{n_2}; \rho_1\cdots \rho_{n_3}; \cdots;     \la_1\cdots \la_{n_{[s]}}}=|x\hb^{\mu_1\cdots \mu_{n_1};\nu_1\cdots \nu_{n_2}; \rho_1\cdots \rho_{n_3}; \cdots;     \la_1\cdots \la_{n_{[s]}}}|\al\hb,\nn\\
\e{1084}
where $|x\hb$ is an eigenstate to the operator $x$.

The quantum theory of the $O(N)$ extended supersymmetric particle model is determined by the conditions \rl{1031} with the above ans\"atze for the states. After replacing $\hat{\psi}$-operators by $b$-operators according to \rl{104} and \rl{105} $\hat{\chi}_a$ is  naturally decomposed as follows
\be
&&\hat{\chi}_a=\Big\{p^2, d_r, d^{\dag}_r, T_{rq}, T^{\dag}_{rq}, I_{rq}\Big\},\quad r,q=1,\ldots,s,\nn\\
\e{109}
for integer spins $s$ (even $N$), and
\be
&&\hat{\chi}_a=\Big\{p^2, d_r, d^{\dag}_r, T_{rq}, T^{\dag}_{rq}, I_{rq}, \cp, \tau_r, \tau^{\dag}_r\Big\},\quad r,q=1,\ldots,[s],\nn\\
&&
\e{110}
for half-integer spins $s$ (odd $N$), where the operators in the curly brackets are defined as follows (I use a different notation here as compared to \cite{Marnelius:1988ab})
\be
&&d_r\equiv p\cdot b_r, \quad T_{rq}\equiv b_r\cdot b_q,\nn\\
&& I_{rq}\equiv \half\Big(b_r^{\dagger}\cdot b_q-b_q\cdot b_r^{\dagger}\Big) = b_r^{\dagger}\cdot b_q-d/2\,\del_{rq}, \nn\\
&&\cp\equiv p\cdot\hps=\cp^{\dag},\quad
\tau_r\equiv \hps\cdot b_r.
\e{111}
Notice that
\be
&&T^{\dag}_{rq}=-b_r^{\dagger}\cdot b_q^{\dagger},\quad T_{rq}=-T_{qr},\nn\\
&&I^{\dag}_{rq}=I_{qr}, \quad \tau^{\dag}_r= -\hps\cdot b^{\dag}_r, \quad d^{\dag}_r=p\cdot b_r^{\dagger}.
\e{1111}
$T$ and $T^{\dag}$ only exist for $s\geq2$ due to their antisymmetry.

The algebra of these operators are (cf.(4.2) and (5.2) in \cite{Marnelius:1988ab}) (the nonzero (anti)commutators)
\be
&&[d_q, d^\dag_r]_+=p^2\del_{qr},\nn\\
&&[T_{qr}, T^\dag_{tu}]_-=\del_{rt}I_{uq}-\del_{qt}I_{ur}+\del_{qu}I_{tr}-\del_{ru}I_{tq},\nn\\
&&[I_{qr}, T_{tu}]_-=\del_{qt}T_{ur}-\del_{qu}T_{tr},\nn\\
&&[I_{qr}, I_{tu}]_-=\del_{rt}I_{qu}-\del_{qu}I_{tr},\nn\\
&&[T_{qr}, d^{\dag}_t]_-=\del_{rt}d_q-\del_{qt}d_r,\nn\\
&&[I_{qr}, d_t]_-=-\del_{qt}d_r,\nn\\
&&[\cp, \cp]_+=p^2,\nn\\
&&[\tau_q, \tau_r]_-=-T_{qr},\nn\\
&&[\tau^{\dag}_q, \tau_r]_-=I_{qr},\nn\\
&&[\tau_q, \cp]_-=d_q,\nn\\
&&[\tau_q, d^{\dag}_r]_-=-\del_{qr}\cp,\nn\\
&&[\tau_q, I_{rt}]_-=\del_{qr}\tau_{t},\nn\\
&&[\tau_q,T^{\dag}_{rt}]_-=\del_{qr}\tau^{\dag}_{t}-\del_{qt}\tau^{\dag}_r.
\e{112}
Notice that $\cp$ and $d_r$ are odd operators. This quantum algebra implies in particular
\be
&&(\cp)^2=\half p^2,\quad (d_r)^2=0\quad r=1,\ldots,[s].
\e{1121}

For integer $s$ (even $N$) the quantum conditions \rl{1031} become (from \rl{109} and where $|F\hb$ is the ansatz \rl{106})
\be
&&(1)\qquad p^2|F\hb=0,\nn\\
&&(2)\qquad d_r|F\hb=0,\nn\\
&&(3)\qquad d_r^{\dagger}|F\hb=0,\nn\\
&&(4)\qquad T_{rq}|F\hb=0,\nn\\
&&(5)\qquad T^{\dagger}_{rq}|F\hb=0,\nn\\
&&(6)\qquad I_{rq}|F\hb=0,
\e{113}
for $r,q=1,\ldots,s$. Condition (1) is the Klein-Gordon equation, (2) and (3) are differential conditions, (4) and (5) are trace conditions, and (6) is an index structure condition. The notation is chosen in accordance with these properties: $d_r$ is a graded external differential operator, $T$ is the trace operator, and $I$ is the index operator. (For $s=0$ only condition (1) exists, and for $s=1$ only (1)-(3) and (6) exist.)

For half-integer $s$ (odd $N$) one gets the  conditions \rl{113} with $|F\hb$ replaced by the general ansatz $|\Psi\hb$ in \rl{108} and for $r,q=1,\ldots [s]$, together with the additional   conditions
\be
&&(7)\qquad \cp|\Psi\hb=0,\nn\\
&&(8)\qquad\tau_r|\Psi\hb=0,\nn\\
&&(9)\qquad\tau_r^{\dagger}|\Psi\hb=0,
\e{114}
for $r=1,\ldots,[s].$
Condition  (7) is the Dirac equation, and (8) and (9) are  gamma trace conditions ($\tau$ is the gamma trace operator).
Notice that the  conditions in \rl{114} imply the conditions in \rl{113} through the algebra \rl{112}. Eqs.\rl{114} are therefore sufficient to determine the properties of the half-integer spin states. (For $[s]=0$ only  condition (7) exists.)

According to the duality properties derived in appendices A and  B also the states $|\tF\hb$ and $|\tPsi\hb$ satisfy \rl{113} and \rl{114} if the states $|F\hb$ and $|\Psi\hb$ do, where  $\tF$ and $\tPsi$ are the dual fields to $F$ and $\Psi$. For this reason it is allowed  to impose (anti)self duality on $F$ and $\Psi$. In addition, the equations \rl{114} are invariant under chiral transformations since $\sqrt{2}\hps^{\mu}$ may be represented by either $\ga^{\mu}$ or $\bar{\ga}\ga^{\mu}$ ($\bar{\ga}^2=-1$) both yielding the same equations. This in turn allows for chiral projections.

It is now straight-forward to write down the corresponding wave equations from the ans\"{a}tze \rl{106} and \rl{108}. For integer spins $s$ the wave function representation of conditions (1)-(6) in \rl{113} is as follows: First the index condition (6) implies that $F$ in \rl{106} is a tensor field of rank $sd/2$ consisting of $s$ groups of antisymmetric sets of $d/2$ indices:
\be
&&F_{\mu_1\cdots\mu_{d/2};\nu_1\cdots\nu_{d/2};\rho_1\cdots\rho_{d/2};\cdots;\la_1\cdots\la_{d/2}}(x),
\e{1141}
\ie the integer $n_j$ for all $j$  in \rl{106} is required  to be $d/2$ which in turn requires $d$ to be even. 
Furthermore, the  condition (6) on the ansatz \rl{106}  implies that $F$ is symmetric under  interchange of any two groups of antisymmetric indices and that it satisfies the properties
\be
&&F_{[\mu_1\cdots\mu_{d/2};\mu_{d/2+1}]\nu_2\cdots\nu_{d/2};\rho_1\cdots\rho_{d/2};\cdots;\la_1\cdots\la_{d/2}}(x)=0.
\e{115}
Obviously, the condition (6) determines the index structure. The remaining conditions in \rl{113} imply
\be
&&(1)\qquad \Box F_{\mu_1\cdots\mu_{d/2};\nu_1\cdots\nu_{d/2};\rho_1\cdots\rho_{d/2};\cdots;\la_1\cdots\la_{d/2}}(x)=0,\nn\\
&&(2)\qquad \dif^{\mu_1}F_{\mu_1\cdots\mu_{d/2};\nu_1\cdots\nu_{d/2};\rho_1\cdots\rho_{d/2};\cdots;\la_1\cdots\la_{d/2}}(x)=0,\nn\\
&&(3)\qquad \dif_{[\mu_1}F_{\mu_2\cdots\mu_{d/2}];\nu_1\cdots\nu_{d/2};\rho_1\cdots\rho_{d/2};\cdots;\la_1\cdots\la_{d/2}}(x)=0,\nn\\
&&(4)\&(5)\qquad \eta^{\mu_1\nu_1}F_{\mu_1\cdots\mu_{d/2};\nu_1\cdots\nu_{d/2};\rho_1\cdots\rho_{d/2};\cdots;\la_1\cdots\la_{d/2}}(x)=0.
\e{116}
According to appendices A and B condition (5) requires the dual of $F$ to be traceless. (Use  the relations \rl{b9} and \rl{b10} in appendix B.) This is satisfied, however,  if $F$ itself is traceless as required by condition (4).

For half-integer spins one finds  wave functions $\Psi$ which are spinors with tensor indices.  The index condition (6) requires the spinor $\Psi$ to  be a tensor field of rank $[s]d/2$ consisting of $[s]$ groups of antisymmetric sets of $d/2$ indices:
\be
&&\Psi_{\mu_1\cdots\mu_{d/2};\nu_1\cdots\nu_{d/2};\rho_1\cdots\rho_{d/2};\cdots;\la_1\cdots\la_{d/2}}(x),
\e{117}
\ie the integer $n_j$ for all $j$  in \rl{108} is required  to be $d/2$ which in turn requires $d$ to be even. 
Furthermore, the  condition (6) in \rl{113} on \rl{108}  implies that $\Psi$ must be symmetric under  interchange of any two groups of antisymmetric indices and that it satisfies the properties (cf \rl{115})
\be
&&\Psi_{[\mu_1\cdots\mu_{d/2};\mu_{d/2+1}]\nu_2\cdots\nu_{d/2};\rho_1\cdots\rho_{d/2};\cdots;\la_1\cdots\la_{d/2}}(x)=0.
\e{118}
The conditions in \rl{114} imply
\be
&&(7)\qquad \cdif\Psi_{\mu_1\cdots\mu_{d/2};\nu_1\cdots\nu_{d/2};\rho_1\cdots\rho_{d/2};\cdots;\la_1\cdots\la_{d/2}}(x)=0,\nn\\
&&(8)\qquad  \ga^{\mu_1}\Psi_{\mu_1\cdots\mu_{d/2};\nu_1\cdots\nu_{d/2};\rho_1\cdots\rho_{d/2};\cdots;\la_1\cdots\la_{d/2}}(x)=0,\nn\\
&&(9)\qquad \ga_{[\mu_1}\Psi_{\mu_2\cdots\mu_{d/2+1}];\nu_1\cdots\nu_{d/2};\rho_1\cdots\rho_{d/2};\cdots;\la_1\cdots\la_{d/2}}(x)=0.
\e{119}
Notice that (9) is the $\ga$-trace condition of the dual of $\Psi$ (see \rl{b9}, \rl{b10} in appendix B).

\setcounter{equation}{0}
\section{Gauge fields and their equations}
\subsection{Definitions of gauge fields}
 From the algebra \rl{112} it is easily seen that condition (3) in \rl{113} is solved by
\be
&&|F\hb=\Pi |\phi\hb,\nn\\
&&|\Psi\hb=\Pi |\rho\hb,
\e{120}
where
\be
&&\Pi\equiv\Pi_{r=1}^{[s]} d^{\dag}_r\equiv d^{\dag}_1d^{\dag}_2\cdots d^{\dag}_{[s]}.
\e{121}
(The $d^{\dag}$-operators anticommute. $\Pi$ is odd for odd $[s]$.)
The states $|\phi\hb$  and $|\rho\hb$ represent  gauge fields. ($|\rho\hb$ is a spinor state.)
The states in \rl{120} satisfy the index condition (6) in \rl{113} if the gauge states satisfy
\be
&& I^{(1)}_{rq}|\phi\hb=0, \quad   I^{(1)}_{rq}|\rho\hb=0, \quad I^{(1)}_{rq}\equiv I_{rq}+\del_{rq}.
\e{122}
These conditions are consistent since it follows from the algebra \rl{112} that $I^{(n)}_{rq}$ defined by
\be
&&I^{(n)}_{rq}\equiv I_{rq}+n\del_{rq}
\e{1221}
for any value of $n$ satisfies
\be
&&[I^{(n)}_{qr}, I^{(n)}_{tu}]_-=\del_{rt}I^{(n)}_{qu}-\del_{qu}I^{(n)}_{tr}.
\e{123}
The algebra \rl{112} makes $|F\hb$ and $|\Psi\hb$ in \rl{120}   invariant under the following  transformations of the gauge states ($|\xi_r\hb$ are spinor states)
\be
&&|\phi\hb\quad\longrightarrow\quad |\phi\hb+\sum_{r=1}^sd_r^{\dagger}|\varepsilon_r\hb,\nn\\
&&|\rho\hb\quad\longrightarrow\quad |\rho\hb+\sum_{r=1}^{[s]}d_r^{\dagger}|\xi_r\hb,
\e{1231}
which also satisfy the conditions \rl{122} provided
\be
&& I^{(1)}_{rq}|\varepsilon_t\hb=-\del_{rt}|\varepsilon_q\hb,\quad I^{(1)}_{rq}|\xi_t\hb=-\del_{rt}|\xi_q\hb.
\e{1232}
These conditions determine the index structure of $|\varepsilon_r\hb$ and $|\xi_r\hb$. Eq.\rl{1231} are gauge transformations and $\varepsilon_r$ and $\xi_r$ are gauge parameters.

I turn now to the 
 corresponding wave function expressions of the above relations. First  the relations \rl{120} may be written as (constant factors and spinor indices are suppressed)
\be
&&F_{\mu_1\cdots\mu_{d/2};\nu_1\cdots\nu_{d/2};\rho_1\cdots\rho_{d/2};\cdots;\la_1\cdots\la_{d/2}}(x)=\nn\\&&=\dif_{[\la_1}\cdots\dif_{[\rho_1}\dif_{[\nu_1}\dif_{[\mu_1}\phi_{\mu_2\cdots\mu_{d/2}];\nu_2\cdots\nu_{d/2}];\rho_2\cdots\rho_{d/2}];\cdots;\la_2\cdots\la_{d/2}]}(x),\nn\\
&&\Psi_{\mu_1\cdots\mu_{d/2};\nu_1\cdots\nu_{d/2};\rho_1\cdots\rho_{d/2};\cdots;\la_1\cdots\la_{d/2}}(x)=\nn\\&&=\dif_{[\la_1}\cdots\dif_{[\rho_1}\dif_{[\nu_1}\dif_{[\mu_1}\rho_{\mu_2\cdots\mu_{d/2}];\nu_2\cdots\nu_{d/2}];\rho_2\cdots\rho_{d/2}];\cdots;\la_2\cdots\la_{d/2}]}(x),\nn\\
\e{124}
where $\phi$ and $\rho$ are of rank $(d/2-1)[s]$. They are antisymmetric within each index block $\{\mu_k\}_{k=1}^{d/2-1},\{\nu_k\}_{k=1}^{d/2-1},\{\rho_k\}_{k=1}^{d/2-1}\cdots$, and symmetric under interchange of any two of these antisymmetric blocks. They also  satisfy the identities
\be
&&\phi_{[\mu_1\cdots\mu_{d/2};\nu_1]\nu_2\cdots\nu_{d/2};\rho_1\cdots\rho_{d/2};\cdots;\la_1\cdots\la_{d/2}}(x)=0,\nn\\
&&\rho_{[\mu_1\cdots\mu_{d/2};\nu_1]\nu_2\cdots\nu_{d/2};\rho_1\cdots\rho_{d/2};\cdots;\la_1\cdots\la_{d/2}}(x)=0,
\e{125}
from \rl{122} above.
 The expression \rl{124} are invariant under the gauge transformations
\be
&&\phi_{\mu_1\cdots\mu_{d/2};\nu_1\cdots\nu_{d/2};\rho_1\cdots\rho_{d/2};\cdots;\la_1\cdots\la_{d/2}}(x)\;\;\longrightarrow\;\;\nn\\&&\phi_{\mu_1\cdots\mu_{d/2};\nu_1\cdots\nu_{d/2};\rho_1\cdots\rho_{d/2};\cdots;\la_1\cdots\la_{d/2}}(x)+\nn\\&&+\sum_{sym\;1,\ldots,s}\dif_{[\mu_1}\varepsilon_{\mu_2\cdots\mu_{d/2}];\nu_1\cdots\nu_{d/2};\rho_1\cdots\rho_{d/2};\cdots;\la_1\cdots\la_{d/2}}(x),\nn\\
&&\rho_{\mu_1\cdots\mu_{d/2};\nu_1\cdots\nu_{d/2};\rho_1\cdots\rho_{d/2};\cdots;\la_1\cdots\la_{d/2}}(x)\;\;\longrightarrow\;\;\nn\\&&\rho_{\mu_1\cdots\mu_{d/2};\nu_1\cdots\nu_{d/2};\rho_1\cdots\rho_{d/2};\cdots;\la_1\cdots\la_{d/2}}(x)+\nn\\&&+\sum_{sym\;1,\ldots,[s]}\dif_{[\mu_1}\xi_{\mu_2\cdots\mu_{d/2}];\nu_1\cdots\nu_{d/2};\rho_1\cdots\rho_{d/2};\cdots;\la_1\cdots\la_{d/2}}(x),
\e{126}
from \rl{1231} above. $\Psi$, $\rho$ and $\xi$ are spinors.
The (vector) index structure of the functions $\varepsilon$ and $\xi$ are determined by \rl{1232}. They are antisymmetric in each numbered block and symmetric under interchange of any two blocks. For integer spins these relations are well-known and were \eg given in \cite{Marnelius:2008er}.

In the following I will not write down the corresponding field expressions with all the indices. The above examples should be a sufficient guide how to do it.

\subsection{Definitions in terms  of dual gauge fields}
It is clear that instead of solving condition (3) in \rl{113} by \rl{120} one may solve condition (2) by
\be
&&|F\hb=\Pi^{\dag} |\Phi\hb,\nn\\
&&|\Psi\hb=\Pi^{\dag} |\cro\hb,
\e{1271}
where (see \rl{121})
\be
&&\Pi^{\dag}=\Pi_{r=1}^{[s]} d_r\equiv d_{[s]}d_{[s]-1}\cdots d_2d_1.
\e{1272}
The states $|\Phi\hb$  and $|\cro\hb$ (a spinor state) represent  new gauge states which have  the dual properties to the previous ones in \rl{120} (see below). The states in \rl{1271} satisfy the index condition (6) in \rl{113} provided these gauge states satisfy
\be
&& I^{(-1)}_{rq}|\Phi\hb=0, \quad   I^{(-1)}_{rq}|\cro\hb=0, \quad I^{(-1)}_{rq}\equiv I_{rq}-\del_{rq}.
\e{1273}
which are consistent due to \rl{123}.
$|F\hb$ and $|\Psi\hb$ in \rl{1271} are obviously  invariant under the following transformations of the dual gauge states
\be
&&|\Phi\hb\quad\longrightarrow\quad |\Phi\hb+\sum_{r=1}^sd_r|\cE_r\hb,\nn\\
&&|\cro\hb\quad\longrightarrow\quad |\cro\hb+\sum_{r=1}^{[s]}d_r|\cX_r\hb,
\e{1275}
which satisfy the conditions \rl{1273} provided
\be
&& I^{(-1)}_{rq}|\cE_t\hb=-\del_{qt}|\cE_r\hb,\quad I^{(-1)}_{rq}|\cX_t\hb=-\del_{qt}|\cX_r\hb,
\e{1276}
which determines the index structure of $|\cE_t\hb$ and $|\cX_t\hb$. 

In fact, the present solutions of conditions (2) and (6) in \rl{113} have dual properties compared  to the previous solutions of conditions (3) and (6) in \rl{113}. To be more precise I define duality  here  by the exchange $b^{\dag\mu}_s\longleftrightarrow b_s^{\mu}$ together with a corresponding exchange of the vacuum states for the $b$-operators. The relation between the vacuum states are given in appendix A and the general duality properties are written down in appendix B. These properties applied to the equations (1)-(9) in \rl{113} and \rl{114} imply   that the dual fields $\tilde{F}$ and $\tilde{\Psi}$ satisfy the same equations as $F$ and $\Psi$. It follows then that one may impose (anti)self-duality on the fields $F$ and $\Psi$. 
From appendix B it follows also  that \rl{120} implies
\be
&&|\tF\hb=\tPi |\tphi\hb,\quad
|\tPsi\hb=\tPi |\trho\hb,
\e{12761}
and since the dual of $\Pi$ is defined to be $\Pi$ with $b$  and $b^{\dag}$ interchanged one has
\be
&&\tPi=(-1)^{{{[s]([s]-1)}\over 2}}\Pi^{\dag}.
\e{1277}
Thus, apart from signs the gauge fields $\Phi$ and $\cR$ in \rl{1271} may  be identified with the dual gauge fields to $\phi$ and $\rho$ in subsection 3.1 in the case of (anti) self-duality.  Also $\cE$ and $\cX$ are then duals to $\varepsilon$ and $\xi$. Thus, in this case one may make  the following identifications
\be
&&\Phi=\tphi,\quad\cro=\trho,\quad\cE_r=\tve_r,\quad\cX_r=\txi_r,
\e{1274}
disregarding signs. By inspection one finds then that \rl{1273}, \rl{1275} and \rl{1276} are  dual relations to \rl{122}, \rl{1231} and \rl{1232} according to the general duality properties in appendix B.

\subsection{General equations for the gauge fields}
In  subsection 3.1 the conditions (3) and (6) in \rl{113} were solved by expressing the original fields in terms of gauge fields. Consider now condition (4), the trace condition. For integer spins I find from \rl{120}
\be
&&T_{rq}|F\hb=\Pi''_{rq}T_{rq}d^{\dag}_rd^{\dag}_q|\phi\hb=-\Pi''_{rq}|E_{\phi,\la}^{(rq)}\hb,\quad |E_{\phi,\la}^{(rq)}\hb\equiv p^2|\phi\hb-d^{\dag}_rd_r|\phi\hb-\nn\\&&-d^{\dag}_qd_q|\phi\hb-d^{\dag}_rd^{\dag}_qT_{rq}|\phi\hb+\sum_{k\neq r,q}d^{\dag}_k|\la_k\hb,
\e{128}
where $\Pi''_{rq}$ is equal to the operator $\Pi$ in \rl{121} without the factors $d^{\dag}_r$ and $d^{\dag}_q$. The sign is defined by the equality $\Pi=\Pi''_{rq}d^{\dag}_rd^{\dag}_q$.  $|\la_k\hb$ are arbitrary states.
It follows that 
\be
&&T_{rq}|F\hb=-\Pi''_{rq}|E_{\phi,\la}\hb,
\e{1282}
where
\be
&& |E_{\phi,\la}\hb\equiv p^2|\phi\hb-\sum_{r=1}^sd^{\dag}_rd_r|\phi\hb-\nn\\&&
-\half \sum_{r,q=1}^sd^{\dag}_rd^{\dag}_qT_{rq}|\phi\hb+{1\over6}\sum_{r,q,t=1}^sd^{\dag}_rd^{\dag}_qd^{\dag}_t|\la_{rqt}\hb,
\e{1281}
where $|\la_{rqt}\hb$ is an arbitrary antisymmetric state.
Since the equality  \rl{1282} is valid for arbitrary $r$ and $q$, condition (4) in \rl{113} is  solved by
\be
&&|E_{\phi,\la}\hb=0,
\e{129}
which is a second order equation for $|\phi\hb$.  (In fact, the corresponding manifestly conformal theory  requires  that $|\phi\hb$ satisfies a second order equation \cite{Arvidsson:2006}.) 
 In order for $|E_{\phi,\la}\hb$ to satisfy
\be
&& I^{(1)}_{rq}|E_{\phi,\la}\hb=0
\e{130}
as required by \rl{122}, $|\la_{rqt}\hb$ must have the index structure determined by
\be
&& I^{(1)}_{rq}|\la_{tuv}\hb=-\del_{rt}|\la_{quv}\hb-\del_{ru}|\la_{tqv}\hb-\del_{rv}|\la_{tuq}\hb.
\e{131}
Since \rl{120} also implies
\be
&&d_r|F\hb=(-1)^{s-1}\Pi'_rd_rd_r^{\dag}|\phi\hb=(-1)^{s-1}\Pi'_r\Big(p^2-d^{\dag}_rd_r\Big)|\phi\hb=\nn\\&&=(-1)^{s-1}\Pi'_r|E_{\phi,\la}\hb,
\e{132}
it follows that condition (2) in \rl{113} is also solved by \rl{129}. ($\Pi'_r$ is $\Pi$ in \rl{121} without the factor $d^{\dag}_r$ and where the   sign is determined by the equality $\Pi=\Pi'_rd^{\dag}_r$.)  Obviously, also (1) is solved by \rl{129} since 
\be
&&p^2|F\hb=\Pi p^2|\phi\hb=\Pi|E_{\phi,\la}\hb.
\e{1321}
In fact, even condition (5) in \rl{113} is solved by \rl{129}.   To see this one may notice that  (see appendix B)
\be
&&T^{\dag}_{rq}|F\hb=0\quad\Leftrightarrow\quad T_{rq}|\tilde{F}\hb=0, 
\e{1322}
and since the trace of the dual of $F$ is zero when the trace of $F$ is zero it follows that condition (5) is satisfied by \rl{129}.   Thus, conditions (1)-(6) are solved by the equation \rl{129} provided the gauge field $\phi$ is defined by \rl{120} and has the index structure \rl{122}. 

The conditions (1)-(6) in \rl{113} are gauge invariant. Also \rl{129} is gauge invariant  if the states $|\la_{rqt}\hb$  transform in a particular way. To see this let the gauge transformation \rl{1231} be combined with
\be
&&|\la_{rqt}\hb \quad\longrightarrow\quad  |\la_{rqt}\hb+\del|\la_{rqt}\hb.
\e{133}
Then a gauge transformation of $|E_{\phi,\la}\hb$ is 
\be
&&|E_{\phi,\la}\hb \quad\longrightarrow\quad  |E_{\phi,\la}\hb+\del|E_{\phi,\la}\hb,\nn\\&& \del|E_{\phi,\la}\hb\equiv -\half\sum_{rqt=1}^sd^{\dag}_rd^{\dag}_qd^{\dag}_t T_{rq}|\varepsilon_t\hb+{1\over6}\sum_{r,q,t=1}^sd^{\dag}_rd^{\dag}_qd^{\dag}_t\del|\la_{rqt}\hb.\nn\\
\e{134}
Obviously $\del|E_{\phi,\la}\hb=0$ if
\be
&&\del|\la_{rqt}\hb=T_{rq}|\varepsilon_t\hb+cycle(rqt).
\e{135}
(For $|\la_{rqt}\hb=0$  \rl{129} reduces to  the (generalized) Fronsdal equations \cite{Fronsdal:1978rb}. See subsection 4.2 below.)

For half-integer spins the decomposition \rl{120} yields
\be
&&\tau_{r}|\Psi\hb=\Pi'_{r}\tau_rd^{\dag}_r|\rho\hb=-\Pi'_{r}|\cE^{(r)}_{\rho,\la}\hb,\nn\\&& |\cE^{(r)}_{\rho,\la}\hb\equiv \cp|\rho\hb-d^{\dag}_r\tau_r|\rho\hb+\sum_{k\neq r}d^{\dag}_k|\la_k\hb,
\e{136}
where $\Pi'_{r}$ as above is equal to the operator $\Pi$ in \rl{121} without the factor $d^{\dag}_r$ and where the  sign is determined by the equality $\Pi=\Pi'_rd^{\dag}_r$. $|\la_k\hb$ are arbitrary spinor states.
It follows that 
\be
&&\tau_{r}|\Psi\hb=-\Pi'_{r}|\cE_{\rho,\la}\hb,
\e{1362}
where
\be
&&|\cE_{\rho,\la}\hb\equiv \cp|\rho\hb-\sum_{r=1}^{[s]}d^{\dag}_r\tau_r|\rho\hb+\half\sum_{r,q=1}^{[s]}d^{\dag}_rd^{\dag}_q|\la_{rq}\hb,
\e{1361}
for arbitrary $r$ where $|\la_{rq}\hb$ is an arbitrary antisymmetric state. 
Obviously
condition (8) in \rl{114} is solved by
\be
&&|\cE_{\rho,\la}\hb=0.
\e{137}
In order for $|\cE_{\rho,\la}\hb$ to satisfy the index condition
\be
&& I^{(1)}_{rq}|\cE_{\rho,\la}\hb=0
\e{142}
as may be inferred from \rl{122}, $|\la_{rq}\hb$ must satisfy
\be
&& I^{(1)}_{rq}|\la_{tu}\hb=-\del_{rt}|\la_{qu}\hb+\del_{ru}|\la_{qt}\hb,
\e{143}
which determines the index structure of $|\la_{rq}\hb$.
Since
\be
&&\cp|\Psi\hb=(-1)^{[s]}\Pi\cp|\rho\hb=(-1)^{[s]}\Pi|\cE_{\rho,\la}\hb,
\e{138}
it follows that \rl{137} also solves condition (7) in \rl{114}.  Also condition (9) in  \rl{114}  should be  solved by \rl{137}, but I have no proof.  Notice that the property
\be
&&\tau^{\dag}_r|\Psi\hb=0\quad\Leftrightarrow\quad   \tau_r|\tPsi\hb=0
\e{1381}
has no consequence here other than that condition (9) is obviously satisfied by \rl{137} if (anti)self-duality is imposed on $\Psi$.
If conditions (7)-(9) are solved by \rl{137}  then  it follows by consistency that also (1)-(6) are solved by \rl{137}. Now, conditions (1)-(6) are actually solved by \rl{137}. To see this one may notice that the above analysis implies that conditions  (1)-(6) are solved by
\be
&&|E_{\rho,\la'}\hb=0,
\e{1382}
where $|E_{\rho,\la'}\hb$ has the form \rl{1281} with $\phi$ and $\la$ replaced by $\rho$ and $\la'$ both with a spinor index. The expression \rl{1361} implies on the other hand 
\be
&&\Big(2\cp-\sum_{r=1}^{[s]}d^{\dag}_r\tau_r\Big)|\cE_{\rho,\la}\hb=\Big(p^2-\sum_{r=1}^{[s]}d^{\dag}_rd_r
-\half \sum_{r,q=1}^{[s]}d^{\dag}_rd^{\dag}_qT_{rq}\Big)|\rho\hb+\nn\\&&+{1\over6}\sum_{r,q,t=1}^{[s]}d^{\dag}_rd^{\dag}_qd^{\dag}_t|\la'_{rqt}\hb,
\e{144}
where $\la'$ is 
\be
&&|\la'_{rqt}\hb\equiv-\Big(\tau_r|\la_{qt}\hb+\tau_t|\la_{rq}\hb+\tau_q|\la_{tr}\hb\Big).
\e{1441}
If $\la'$ in \rl{1382} is chosen to be this expression then \rl{144} is equal to $|E_{\rho,\la'}\hb$ and \rl{1382} follows from \rl{137}.
 (Equation \rl{1382} is except for the spinor part the equation for a spin $[s]=s-1/2$ particle.)
In conclusion, conditions (1)-(8) are  solved by \rl{137}, and probably condition (9) as well. (It is at least solved if  (anti)self-duality is imposed on $\Psi$.)  All this provided the gauge fields $\rho$ have the index structure \rl{122}.

The gauge transformation
\be
&&|\rho\hb \quad\longrightarrow\quad |\rho\hb   +\sum_{k=1}^{[s]}d^{\dag}_k|\xi_k\hb
\e{139}
implies that 
$|\cE_{\rho,\la}\hb$ in \rl{1361} transforms as follows
\be
&&\del|\cE_{\rho,\la}\hb=-\sum_{r,q=1}^{[s]}d^{\dag}_rd^{\dag}_q\tau_r|\xi_q\hb+\half\sum_{r,q=1}^{[s]}d^{\dag}_rd^{\dag}_q\del|\la_{rq}\hb,
\e{140}
where $\del|\la_{rq}\hb$ is the gauge transformation of $|\la_{rq}\hb$. $|\cE_{\rho,\la}\hb$ is therefore gauge invariant if
\be
&&\del|\la_{rq}\hb=\tau_r|\xi_q\hb-\tau_q|\xi_r\hb.
\e{141}
(For $|\la_{rq}\hb=0$ \rl{137} reduces to the (generalized) Fang-Fronsdal equations \cite{Fang:1978wz}. See subsection 4.2.) Notice that \rl{141} implies for \rl{1441}
\be
&&\del|\la'_{rqt}\hb=T_{rq}|\xi_t\hb+cycle(rqt),
\e{1442}
which makes \rl{1382} gauge invariant (cf \rl{135}) as is obvious from \rl{144}.

The derivations  in \rl{1282} and \rl{1362} are similar to previous derivations (see \eg \cite{Labastida:1986ft,Labastida:1986zb,Bekaert:2002dt,Bandos: 2005mb,Bekaert:2003az,Bekaert:2006ix}), and in fact they were inspired by  the derivations  in \cite{Bandos: 2005mb} where the same field  representation is used.

\subsection{General equations for the dual gauge fields}
In subsection 3.2 the original gauge invariant fields were expressed in terms of dual gauge fields when I solved conditions (2) and (6) in \rl{113}. Consider now the dual trace condition (5). For integer spins I find from \rl{1271}
\be
&&T^{\dag}_{rq}|F\hb=-(\Pi''_{rq})^{\dag}T^{\dag}_{rq}d_rd_q|\Phi\hb=-(\Pi''_{rq})^{\dag}|G_{\Phi,\Lambda}^{(rq)}\hb,\nn\\
&& |G_{\Phi,\Lambda}^{(rq)}\hb\equiv p^2|\Phi\hb-d_rd^{\dag}_r|\Phi\hb-d_qd^{\dag}_q|\Phi\hb-d_qd_rT^{\dag}_{rq}|\Phi\hb+\sum_{k\neq r,q}d_k|\Lambda_k\hb,\nn\\
\e{145}
where $\Pi^{\dag}=-(\Pi''_{rq})^{\dag}d_rd_q$. ($(\Pi''_{rq})^{\dag}$ is equal to the operator $\Pi^{\dag}$ in \rl{1272} without the factors $d_r$ and $d_q$.)
It follows that 
\be
&&T^{\dag}_{rq}|F\hb=-(\Pi''_{rq})^{\dag}|G_{\Phi,\Lambda}\hb,
\e{1451}
where
\be
&&|G_{\Phi,\Lambda}\hb\equiv p^2|\Phi\hb-\sum_{r=1}^sd_rd^{\dag}_r|\Phi\hb-\half \sum_{r,q=1}^sd_qd_rT^{\dag}_{rq}|\Phi\hb+\nn\\&&\qquad\qquad\qquad+{1\over6}\sum_{r,q,t=1}^sd_rd_qd_t|\Lambda_{rqt}\hb,
\e{146}
where $|\Lambda_{rqt}\hb$ is an arbitrary antisymmetric state. Thus, 
condition (5) is solved by
\be
&&|G_{\Phi,\Lambda}\hb=0.
\e{1461}
In order for $|G_{\Phi,\Lambda}\hb$ to satisfy
\be
&& I^{(-1)}_{rq}|G_{\Phi,\Lambda}\hb=0
\e{147}
as may be inferred from \rl{1273}, $|\Lambda_{rqt}\hb$ must satisfy
\be
&& I^{(-1)}_{rq}|\Lambda_{tuv}\hb=\del_{qt}|\Lambda_{ruv}\hb+\del_{qu}|\Lambda_{trv}\hb+\del_{qv}|\Lambda_{tur}\hb,
\e{148}
where $I^{(-1)}$ is defined in \rl{1221}. These conditions determine the index structure of $\Lambda_{tuv}$.
Since
\be
&&d^{\dag}_r|F\hb=(-1)^{s-1}(\Pi'_r)^{\dag}d_r^{\dag}d_r|\Phi\hb=(-1)^{s-1}(\Pi'_r)^{\dag}\Big(p^2-d_rd^{\dag}_r\Big)|\Phi\hb=\nn\\&&=(-1)^{s-1}(\Pi'_r)^{\dag}|G_{\Phi,\Lambda}\hb,
\e{149}
it follows that condition (3) in \rl{113} is also solved by \rl{1461}. ($(\Pi)^{\dag}=(-1)^{[s]-1}(\Pi'_r)^\dag d_r$. Thus, $(\Pi'_r)^{\dag}$ is $\Pi^{\dag}$ in \rl{1272} without the factor $d_r$.) Condition (1) is solved by \rl{1461} since
\be
&&p^2|F\hb=\Pi^{\dag}p^2|\Phi\hb=\Pi^{\dag}|G_{\Phi,\Lambda}\hb.
\e{1491}
Also condition (4) is solved by \rl{146} since the equivalence \rl{1322} is still valid.
Thus, conditions (1)-(6) are solved by the equation \rl{146} provided the gauge field $\Phi$ is defined by \rl{1271} and has the index structure \rl{1273}.

The conditions (1)-(6)  in \rl{113} are gauge invariant. Also \rl{146} is gauge invariant  if the states $|\Lambda_{rqt}\hb$  transform in a particular way. Let the gauge transformation \rl{1275} be combined with
\be
&&|\Lambda_{rqt}\hb \quad\longrightarrow\quad  |\Lambda_{rqt}\hb+\del|\Lambda_{rqt}\hb.
\e{150}
Then a gauge transformation of $|G_{\Phi,\Lambda}\hb$ is 
\be
&&|G_{\Phi,\Lambda}\hb \quad\longrightarrow\quad  |G_{\Phi,\Lambda}\hb+\del|G_{\Phi,\Lambda}\hb,\nn\\&& \del|G_{\Phi,\Lambda}\hb\equiv -\half\sum_{rqt=1}^sd_rd_qd_t T^{\dag}_{qr}|\cE_t\hb+{1\over6}\sum_{r,q,t=1}^sd_rd_qd_t\del|\Lambda_{rqt}\hb.\nn\\
\e{151}
Obviously $\del|G_{\Phi,\Lambda}\hb=0$ if
\be
&&\del|\Lambda_{rqt}\hb=-T^{\dag}_{rq}|\Lambda_t\hb+cycle(rqt).
\e{152}
 
For half integer spins the decomposition \rl{1271} yields
\be
&&\tau^{\dag}_{r}|\Psi\hb=(-1)^{[s]-1}(\Pi'_{r})^{\dag}\tau^{\dag}_rd_r|\cR\hb=(-1)^{[s]-1}(\Pi'_{r})^{\dag}|\cG^{(r)}_{\cR,\Lambda}\hb,\nn\\&& |\cG^{(r)}_{\cR,\Lambda}\hb\equiv \cp|\cR\hb+d_r\tau^{\dag}_r|\cR\hb+\sum_{k\neq r}d_k|\Lambda_k\hb,
\e{153}
where $(\Pi'_{r})^{\dag}$ is defined as above, \ie it    is equal to the operator $\Pi^{\dag}$ in \rl{1272} without the factor $d_r$.
It follows that 
\be
&&\tau^{\dag}_{r}|\Psi\hb=(-1)^{[s]-1}(\Pi'_{r})^{\dag}|\cG_{\cR,\Lambda}\hb,
\e{1531}
where
\be
&&|\cG_{\cR,\Lambda}\hb\equiv \cp|\cR\hb+\sum_{r=1}^{[s]}d_r\tau^{\dag}_r|\cR\hb+\half\sum_{r,q=1}^{[s]}d_rd_q|\Lambda_{rq}\hb,
\e{154}
where $|\Lambda_{rq}\hb$ is an arbitrary antisymmetric spinor state. Thus, condition (9) in \rl{114} is solved by
\be
&&|\cG_{\cR,\Lambda}\hb=0.
\e{1541}
In order for $|\cG_{\cR,\Lambda}\hb$ to satisfy the index condition
\be
&& I^{(-1)}_{rq}|\cG_{\cR,\Lambda}\hb=0
\e{159}
as may be inferred from \rl{1273}, $|\Lambda_{rq}\hb$ must satisfy
\be
&& I^{(-1)}_{rq}|\Lambda_{tu}\hb=\del_{qt}|\Lambda_{ru}\hb-\del_{qu}|\Lambda_{rt}\hb,
\e{160}
which determines the index structure of $|\Lambda_{rq}\hb$. 
Since
\be
&&\cp|\Psi\hb=(-1)^{[s]}\Pi^{\dag}\cp|\cR\hb=(-1)^{[s]}\Pi^{\dag}|\cG_{\cR,\Lambda}\hb,
\e{155}
it follows that \rl{1541} also solves condition (7) in \rl{114}. Also  condition (8) should be  solved by \rl{1541} but  I have no proof. It is satisfied if (anti)self-duality is imposed on $\Psi$. Conditions (1)-(6) are solved by \rl{1541}. This follows since (1)-(6) are solved by
\be
&&|G_{\cR,\Lambda'}\hb=0,
\e{1551}
where $|G_{\cR,\Lambda'}\hb$ is equal to \rl{146} with $\Phi$ and $\Lambda$ replaced by $\cR$ and $\Lambda'$ both with the same index structure except that the latter also have a spinor index. Notice that 
\be
&&(2\cp+d_r\tau_r^{\dag})|\cG_{\cR,\Lambda}\hb=\Big(p^2-\sum_{r=1}^{[s]}d_rd_r^{\dag}+\half \sum_{r,q=1}^{[s]} d_rd_qT_{rq}^{\dag}\Big)|\cR\hb+\nn\\
&&+{1\over6}\sum_{r,q,t=1}^{[s]}|\Lambda'_{rqt}\hb,
\e{1552}
where
\be
&&|\Lambda'_{rqt}\hb=\tau^{\dag}_r|\Lambda_{qt}\hb+\tau^{\dag}_q|\Lambda_{tr}\hb+\tau^{\dag}_t|\Lambda_{rq}\hb.
\e{1553}
If $\Lambda'$ in  \rl{1551}  is chosen to be the expression  \rl{1553} then \rl{1551} follows from \rl{1541} since  \rl{1552} then is equal to $|G_{\cR,\Lambda'}\hb$ in \rl{1551}.

The gauge transformation
\be
&&|\cR\hb \quad\longrightarrow\quad |\cR\hb   +\sum_{k=1}^{[s]}d_k|\cX_k\hb
\e{156}
makes
$|\cG_{\cR,\Lambda}\hb$ in \rl{154} to transform as follows
\be
&&\del|\cE_{\cR,\Lambda}\hb=\sum_{r,q=1}^{[s]}d_rd_q\tau^{\dag}_r|\cX_q\hb+\half\sum_{r,q=1}^{[s]}d_rd_q\del|\Lambda_{rq}\hb,
\e{157}
where $\del|\Lambda_{rq}\hb$ is the gauge transformation of $|\Lambda_{rq}\hb$. $|\cG_{\cR,\Lambda}\hb$ is therefore gauge invariant if
\be
&&\del|\Lambda_{rq}\hb=\tau^{\dag}_q|\cX_r\hb-\tau^{\dag}_r|\cX_q\hb.
\e{158}
These transformations imply that $\Lambda'$ in \rl{1553} transform as \rl{152}  with $|\Lambda_t\hb$ replaced by $|\cX_t\hb$.

According to the duality rules in appendix B one has the duality properties
\be
&&\widetilde{|G_{\Phi,\Lambda}\hb}=|E_{\tilde{\Phi},\tilde{\Lambda}}\hb,\quad\widetilde{|\cG_{\cR,{\Lambda}}\hb}=|\cE_{\tilde{\cR},\tilde{\Lambda}}\hb,
\e{1601}
where $\tilde{\Phi}$, $\tilde{\cR}$ and $\tilde{\Lambda}$ have the same index structure as $\phi$, $\rho$ and $\la$. Thus, the solutions here are equivalent to the  solutions in the previous subsection. They are identical if (anti)self-duality is imposed on $F$ and $\Psi$. 
If (up to signs), as required by (anti)self-duality,
\be
&&\Phi=\tilde{\phi},\quad\Lambda_{rqt}=\tilde{\la}_{rqt},\quad\cR=\tilde{\rho}, \quad \Lambda_{rq}=\tilde{\la}_{rq},
\e{161}
then  one finds by inspection that  \rl{146}, \rl{147}, \rl{148}, \rl{151}, \rl{152}, \rl{154}, \rl{1552}, \rl{1553}, \rl{159}, \rl{160}, \rl{156}, \rl{157}, \rl{158}  are the dual relations  to  \rl{1281}, \rl{130}, \rl{131}, \rl{134}, \rl{135}, \rl{1361}, \rl{144}, \rl{1441}, \rl{142}, \rl{143}, \rl{139}, \rl{140}, \rl{141}  respectively.

\setcounter{equation}{0}
\section{Properties of the equations for the gauge fields}
 From now on and until the end of part I repeated indices are summed over.
 
\subsection{Identities and auxiliary conditions}
From the expressions \rl{1281} and \rl{1361} for
 $|E_{\phi,\la}\hb$ and $|\cE_{\rho,\la}\hb$ in the equations of motion \rl{129} and \rl{137} one may derive the following identities.
 \be
&&\Big(d_r-\half d_t^{\dag}T_{tr}\Big)|E_{\phi,\la}\hb\equiv{1\over12}d_k^{\dag}d_q^{\dag}d_t^{\dag}|C_{kqtr}\hb,
\e{1611}
\be
&&\Big(d_r+\cp\tau_r-{1\over3}d^{\dag}_t(T_{tr}+\tau_r\tau_t)\Big)|\cE_{\rho,\la}\hb\equiv-{1\over6}d_t^{\dag}d_k^{\dag}|Z_{tkr}\hb,
\e{162}
where
\be
&|C_{kqtr}\hb\equiv&\Big(T_{kq}T_{tr}+T_{qt}T_{kr}+T_{tk}T_{qr}\Big)|\phi\hb-\nn\\&&-d_k|\la_{qtr}\hb+d_q|\la_{trk}\hb-d_t|\la_{rkq}\hb+d_r|\la_{kqt}\hb-\nn\\&&-{1\over2}d_u^{\dag}\Big(T_{kr}|\la_{uqt}\hb+T_{qr}|\la_{utk}\hb+T_{tr}|\la_{ukq}\hb\Big)+\nn\\&&+{1\over2}d_u^{\dag}\Big(T_{kq}|\la_{urt}\hb+T_{qt}|\la_{urk}\hb+T_{tk}|\la_{urq}\hb\Big)+\nn\\&&+{1\over2}d_u^{\dag}\Big(T_{uk}|\la_{qtr}\hb-T_{uq}|\la_{trk}\hb+T_{ut}|\la_{rkq}\hb-T_{ur}|\la_{kqt}\hb\Big),\nn\\
\e{1621}
\be
&|Z_{tkr}\hb\equiv&\Big(T_{kr}\tau_t+T_{rt}\tau_k+T_{tk}\tau_r\Big)|\rho\hb+\Big(d_k|\la_{rt}\hb+d_r|\la_{tk}\hb+d_t|\la_{kr}\hb\Big)-\nn\\&&-\cp\Big(\tau_k|\la_{rt}\hb+\tau_r|\la_{tk}\hb+\tau_t|\la_{kr}\hb\Big)+\nn\\&&+d^{\dag}_q\Big(T_{kr}|\la_{qt}\hb+T_{rt}|\la_{qk}\hb+T_{tk}|\la_{qr}\hb\Big)+\nn\\&&+d_q^{\dag}\tau_q\Big(\tau_k|\la_{rt}\hb+\tau_r|\la_{tk}\hb+\tau_t|\la_{kr}\hb\Big).
\e{163}
These expressions are totally antisymmetric and also fully gauge invariant. Imposing the equations of motion
\be
&&|E_{\phi,\la}\hb=0,\quad|\cE_{\rho,\la}\hb=0
\e{165}
requires then $|C_{kqtr}\hb$ and $|Z_{tkr}\hb$ to be of the form
\be
&&|C_{kqtr}\hb=d_k^{\dag}d_q^{\dag}|A_{tr}\hb+d^{\dag}_qd^{\dag}_t|A_{kr}\hb+ cycle(qtr),  \nn\\&&|Z_{tkr}\hb=d^{\dag}_td^{\dag}_k|A_r\hb+cycle(tkr),
\e{1651}
where $|A_{tr}\hb=-|A_{rt}\hb$. Since $|C_{kqtr}\hb$ and $|Z_{tkr}\hb$ are gauge invariant also $|A_{tr}\hb$ and $|A_r\hb$ must be gauge invariant. However, there seem to be  no (local) gauge invariant states $|A_{tr}\hb$ and $|A_r\hb$ that are nonzero when \rl{165} are imposed. (There seem also to be no gauge invariant $|A_{tr}\hb$ and $|A_r\hb$ consistent with the conformal properties of   $|C_{kqtr}\hb$ and $|Z_{tkr}\hb$.) Effectively, \rl{165} seems therefore to require
\be
&&|C_{kqtr}\hb=0,\quad|Z_{tkr}\hb=0.
\e{164}

In fact, there are further identities:
\be
&&\Big(T_{kq}T_{tr}+T_{qt}T_{kr}+T_{tk}T_{qr}\Big)|E_{\phi,\la}\hb\equiv2p^2|C_{kqtr}\hb-d_u^{\dag}d_u|C_{kqtr}\hb-\nn\\&&-d_td_u^{\dag}|C_{urqk}\hb+d_rd_u^{\dag}|C_{uqkt}\hb-d_qd_u^{\dag}|C_{uktr}\hb+d_kd_u^{\dag}|C_{utrq}\hb-\nn\\&&-\half d_u^{\dag}d_v^{\dag}T_{uv}|C_{kqtr}\hb,
\e{1641}
\be
&&\Big(T_{kr}\tau_t+T_{rt}\tau_k+T_{tk}\tau_r\Big)|\cE_{\rho,\la}\hb\equiv\Big(2\cp-d^{\dag}_u\tau_u\Big)|Z_{krt}\hb+\half d^{\dag}_u\tau_{[u}|Z_{tkr]}\hb,\nn\\
\e{16411}
where the last antisymmetric expression may be written as follows 
\be
&&\tau_{[u}|Z_{tkr]}\hb\equiv \tau_u|Z_{tkr}\hb-\tau_t|Z_{kru}\hb+\tau_k|Z_{rut}\hb-\tau_r|Z_{utk}\hb.
\e{1642}
The identities \rl{1641} and \rl{16411} restrict the expressions \rl{1651} further. (They are, however,  trivially satisfied for $s<4$ and $s<7/2$ respectively.)

From \rl{144} it follows that the  identities in \rl{1611} and \rl{1641} are also satisfied if $|E_{\phi,\la}\hb$ is replaced by $(2\cp-d_r^{\dag}\tau_r)|\cE_{\rho,\la}\hb$. However, the resulting $|C_{kqtr}\hb$ is then given by \rl{1621} with $|\phi\hb$ replaced by $|\rho\hb$, and with $|\la_{kqt}\hb$ replaced by $|\la'_{kqt}\hb$ given in \rl{1441}. This $|C_{kqtr}\hb$ may also be expressed in terms of $|Z_{krt}\hb$ as follows: $|C_{kqtr}\hb=\half \tau_{[k}|Z_{qtr]}\hb$.

In the present formulation the antisymmetric products in \rl{1621}, \rl{163}, \rl{1641} and \rl{16411}, \ie
\be
&&(TT)_{kqtr}\equiv\Big(T_{kq}T_{tr}+T_{qt}T_{kr}+T_{tk}T_{qr}\Big),\nn\\&&(T\tau)_{krt}\equiv\Big(T_{kr}\tau_t+T_{rt}\tau_k+T_{tk}\tau_r\Big)=(\tau T)_{krt},
\e{1643}
should be interpreted as the double trace operator and the product of the trace and gamma trace operator respectively. 
Notice also the antisymmetric triple gamma trace operator
\be 
&&(\tau\tau\tau)_{krt}\equiv\tau_k\tau_r\tau_t+ \tau_r\tau_t\tau_k+ \tau_t\tau_k\tau_r-\tau_r\tau_k\tau_t - \tau_k\tau_t\tau_r-\tau_t\tau_r\tau_k=\nn\\&&=-(T\tau)_{krt},
\e{1644}
and the antisymmetric quadruple gamma trace operator
\be
&&(\tau\tau\tau\tau)_{utkr}\equiv\tau_{u}(\tau\tau\tau)_{tkr}-\tau_{t}(\tau\tau\tau)_{kru}+\tau_{k}(\tau\tau\tau)_{rut}-\tau_{r}(\tau\tau\tau)_{utk}=\nn\\&&=-2(TT)_{utkr}.
\e{1645}
 
One may notice the following  correspondences between the expressions given here and the expressions given  in the symmetric case in \cite{Francia:2005bu}  (in $d=4$ they should be equivalent). \\ \\
\hspace*{2cm}
$\begin {array}{lll}
|E_{\phi,\la}\hb&\longleftrightarrow&\cal{A}\;{\rm in}\;{\rm (2.4)}\\
|\cE_{\rho,\la}\hb&\longleftrightarrow&\cal{W}\;{\rm in}\;{\rm (2.43)}\\
\rl{1611}&\longleftrightarrow&{\rm (2.6)}\\
\rl{162}&\longleftrightarrow&{\rm (2.45)}\\
|C_{kqtr}\hb\;{\rm in}\;\rl{163}&\longleftrightarrow&\cC\;{\rm in}\;{\rm (2.12)}\\
|Z_{tkr}\hb\;{\rm in}\;\rl{164}&\longleftrightarrow&\cal{Z}\;{\rm in}\;{\rm (2.50)}\\
\rl{1641}&\longleftrightarrow&{\rm (2.15)}\\
\rl{16411}&\longleftrightarrow&{\rm (4.15)}.
\end{array}$\\ \\
The equation numbers on the right-hand side refer to  section 2 in \cite{Francia:2007qt} except for the last one which refers to \cite{Francia:2005bu} (the last identity is not written down in \cite{Francia:2007qt}).

\subsection{Relations to the (Fang)-Fronsdal equations}
The generalized (Fang)-Fronsdal equations follow from \rl{165} when the compensator fields are set to zero, \ie
\be
&&|\la_{kq}\hb=0,\quad |\la_{rqt}\hb=0.
\e{166}
In this case \rl{165} reduces to
\be
&&|F_{\phi}\hb=0,\quad |F_{\phi}\hb\equiv\Big(p^2-d_r^{\dag}d_r-\half d_r^{\dag}d_t^{\dag}T_{rt}\Big)|\phi\hb,\nn\\
&&|\cF_{\rho}\hb=0,\quad |\cF_{\rho}\hb\equiv\Big(\cp-d^{\dag}_r\tau_r\Big)|\rho\hb.
\e{167}
From these expressions and the relation
\be
&&\Big(2\cp-d^{\dag}_k\tau_k\Big)\Big(\cp-d^{\dag}_r\tau_r\Big)=p^2-d_r^{\dag}d_r-\half d_r^{\dag}d_t^{\dag}T_{rt}.
\e{1671}
it follows that  $|\rho\hb$ for spin $s$ satisfies   the same equation as $|\phi\hb$ for spin $s-1/2$. (This is a special case of \rl{144}.) 
From \rl{166} it follows furthermore that the gauge invariance is restricted to those in which the gauge parameters satisfy the trace conditions (for all $q,t,r$)
\be
&&T_{rq}|\varepsilon_t\hb+T_{qt}|\varepsilon_r\hb+T_{tr}|\varepsilon_q\hb=0,\nn\\
&&\tau_r|\xi_t\hb-\tau_t|\xi_r\hb=0
\e{168}
from \rl{135} and \rl{141}. It is easy to see that these restrictions are  trivially satisfied for $s\leq2$ and $s\leq3/2$ respectively, and that they are  nontrivial  for $s\geq3$ in the integer spin case, and for $s\geq5/2$ in the half-odd integer case. 
The restrictions \rl{164} imply also from \rl{1621} and \rl{163} that the gauge fields satisfy the double trace conditions which here have the form (for all $k,q,t,r$)
\be
&&(TT)_{kqtr}|\phi\hb=0,\quad (T\tau)_{krt}|\rho\hb=0,
\e{169}
where the totally antisymmetric  expressions $(TT)$ and $(T\tau)$ are defined in \rl{1643}.
It is easily seen that they are trivially satisfied for $s\leq3$ and $s\leq5/2$ respectively, and   nontrivial  for $s\geq4$  and $s\geq7/2$ respectively.
The above results should be sufficiently convincing that the restricted equations \rl{167} for $d=4$ indeed are equivalent to the (Fang)-Fronsdal equations, and that their restrictions on the gauge parameters and gauge fields are exactly \rl{168} and \rl{169} for $d=4$. (Only in $d=4$ do I have symmetric gauge fields.) For even dimensions $d>4$ \rl{167}-\rl{169} are generalized (Fang)-Fronsdal equations.

Notice that even the restricted case satisfies identities like \rl{1611}, \rl{162} and \rl{1641},\rl{16411}:
\be
&&\Big(d_r-\half d_t^{\dag}T_{tr}\Big)|F_{\phi}\hb={1\over12}d_k^{\dag}d_q^{\dag}d_t^{\dag}(TT)_{kqtr}|\phi\hb,\nn\\
&&\Big(d_r+\cp\tau_r-{1\over3}d_t^{\dag}(T_{tr}+\tau_r\tau_t)\Big)|\cF_{\rho}\hb=-{1\over6}d_t^{\dag}d_k^{\dag}(T\tau)_{krt}|\rho\hb,
\e{1691}
and
\be
&&(TT)_{kqtr}|F_{\phi}\hb\equiv2p^2(TT)_{kqtr}|\phi\hb-d_u^{\dag}d_u(TT)_{kqtr}|\phi\hb-\nn\\&&-d_td_u^{\dag}(TT)_{urqk}|\phi\hb+d_rd_u^{\dag}(TT)_{uqkt}|\phi\hb-\nn\\&&-d_qd_u^{\dag}(TT)_{uktr}|\phi\hb+d_kd_u^{\dag}(TT)_{utrq}|\phi\hb-\nn\\&&-\half d_u^{\dag}d_v^{\dag}T_{uv}(TT)_{kqtr}|\phi\hb,\nn\\
&&(T\tau)_{krt}|\cF_{\rho}\hb\equiv\Big(2\cp-d^{\dag}_u\tau_u\Big)(T\tau)_{krt}|\rho\hb+\half d^{\dag}_u\tau_{[u}(T\tau)_{tkr]}|\rho\hb,
\e{1692}
where  the totally antisymmetric expressions $(TT)$ and $(T\tau)$ are defined in \rl{1643},
and where 
\be
&&\tau_{[u}(T\tau)_{tkr]}\equiv \tau_u(T\tau)_{tkr}-\tau_t(T\tau)_{kru}+\tau_k(T\tau)_{rut}-\tau_r(T\tau)_{utk}=\nn\\
&&\quad\quad\quad\quad\quad =2(TT)_{utkr}.
\e{16931}

The identities \rl{1691} and \rl{1692} follow  from the operator equalities
\be
&&\Big(d_r-\half d_t^{\dag}T_{tr}\Big)\Big(p^2-d^{\dag}_ud_u-\half d^{\dag}_ud^{\dag}_vT_{uv}\Big)={1\over12}d_k^{\dag}d_q^{\dag}d_t^{\dag}(TT)_{kqtr},\nn\\
&&\Big(d_r+\cp\tau_r-{1\over3}d_t^{\dag}(T_{tr}+\tau_r\tau_t)\Big)\Big(\cp-d_u^{\dag}\tau_u\Big)=-{1\over6}d_t^{\dag}d_k^{\dag}(T\tau)_{krt},\nn\\
&&[(TT)_{kqtr}, p^2-d^{\dag}_ud_u-\half d^{\dag}_ud^{\dag}_vT_{uv}]_-=\Big(p^2-d_ud_u^{\dag}\Big)(TT)_{kqtr},\nn\\
&&[(T\tau)_{krt}, \cp-d_u^{\dag}\tau_u]_-=\cp(T\tau)_{krt}+d_u^{\dag}(TT)_{ukrt},
\e{1694}
obtained from the quantum algebra \rl{112}.

\setcounter{equation}{0}
\section{Gauge invariant actions for the wave functions}
I shall now look for  actions which reproduce the gauge invariant equations above. Such actions must be gauge invariant. A natural real ansatz  for the actions of the equations \rl{129} and \rl{137} (or equivalently \rl{165}) are
\be
&&S_{01}\equiv\vb\phi|E_{\phi,\la}\hb+\vb E_{\phi,\la}|\phi\hb,\quad S_{02} \equiv \vb\rho|\cE_{\rho,\la}\hb+\vb \cE_{\rho,\la}|\rho\hb,
\e{170}
where the gauge fields $|\phi\hb$, $|\rho\hb$ only are restricted by their index structure, \ie the conditions in \rl{122} must be satisfied.
Now, there is a problem with the gauge invariance of these actions. 
Although $|E_{\phi,\la}\hb$ and $|\cE_{\rho,\la}\hb$ are gauge invariant, a gauge transformation \rl{1231} of the actions \rl{170} yields
\be
&&S_{01}\quad\longrightarrow\quad S_{01}+\vb\varepsilon_k|d_k|E_{\phi,\la}\hb+\vb E_{\phi,\la}|d_k^{\dag}|\varepsilon_k\hb,\nn\\
&&S_{02} \quad\longrightarrow\quad S_{02}+\vb\xi_k|d_k|\cE_{\rho,\la}\hb+\vb \cE_{\rho,\la}|d_k^{\dag}|\xi_k\hb.
\e{171}
Gauge invariance of the actions \rl{170} requires therefore
\be
&&d_k|E_{\phi,\la}\hb=0,\quad d_k|\cE_{\rho,\la}\hb=0,\quad k=1,\ldots,[s].
\e{172}
However, these conditions are not satisfied for $s>1$. I find
\be
&&d_k |E_{\phi,\la}\hb=\Big(-d^{\dag}_r d_rd_k +p^2d^{\dag}_rT_{rk}-\half  d^{\dag}_r d^{\dag}_qT_{rq}d_k\Big)|\phi\hb+\nn\\
&&\qquad\qquad+\half p^2   d^{\dag}_rd^{\dag}_q|\la_{rqk}\hb-{1\over6}d^{\dag}_r d^{\dag}_qd^{\dag}_td_k|\la_{rqt}\hb,    \nn\\
&&d_k|\cE_{\rho,\la}\hb=-\Big(\cp d_k+p^2\tau_k -d^{\dag}_r\tau_r d_k\Big) |\rho\hb+\nn\\&&\qquad\qquad+p^2d^{\dag}_r |\la_{kr}\hb+\half d^{\dag}_rd^{\dag}_q d_k |\la_{rq}\hb.
\e{173}
Obviously   gauge invariant actions must contain more terms than those in \rl{170}. In fact, since the gauge invariance of $|E_{\phi,\la}\hb$ and $|\cE_{\rho,\la}\hb$ requires gauge transformations of the compensator fields (the $|\la\hb$-states), also the $\la$-fields must be dynamical, \ie they must be either expressed in terms of the gauge fields or they must be treated as  independent auxiliary fields.  Here I choose to treat the compensator fields as independent fields.
Thus, the actions  not only yield the equations \rl{129} and \rl{137} (or equivalently \rl{165}) but also equations that should either determine the compensator fields in terms of the gauge fields or leave them redundant. Inspired and guided by the corresponding construction for symmetric gauge fields by Francia and Sagnotti in \cite{Francia:2005bu} (see also section 2 of \cite{Francia:2007qt}) I was finally able to arrive at the following gauge invariant actions: (I am thankful to  Dario Francia for carefully explaining the derivation in \cite{Francia:2005bu,Francia:2007qt})
\be
&&S_1\equiv \half\vb\phi|\Big(1-{1\over4}T^{\dag}_{ut}T_{ut}\Big)|E_{\phi,\la}\hb+{1\over8}\vb\al_{utr}|T_{ut}d_r|E_{\phi,\la}\hb+\nn\\
&&\quad\quad+{1\over24}\vb\beta_{uqtr}|C_{uqtr}\hb+c.c.,\nn\\
&&S_2\equiv\half\vb\rho|\Big(1-\tau^{\dag}_k\tau_k+{1\over3}\tau^{\dag}_k\tau_q^{\dag}(T_{qk}+\tau_k\tau_q)\Big)|\cE_{\rho,\la}\hb+\nn\\
&&\quad\quad+\half\vb\al_{kr}|\Big(d_r\tau_k+{1\over3}\cp(\tau_k\tau_r+\tau_r\tau_k)\Big)|\cE_{\rho,\la}\hb-\nn\\
&&\quad\quad-{1\over6}\vb\al_{kr}|\tau^{\dag}_q(T_{qk}+\tau_k\tau_q)d_r|\cE_{\rho,\la}\hb-{1\over12}\vb\beta_{tkr}|Z_{tkr}\hb+c.c.,
\e{174}
where I apart from the gauge fields have introduced the independent auxiliary fields $\al$ and $\beta$. They are defined below. 

In the integer spin case $\al$ is required to transform as
\be
\del|\al_{utr}\hb=T_{ut}|\varepsilon_r\hb
\e{175}
 under gauge transformations. The compensator field $\la$ in $|E_{\phi,\la}\hb$ is then defined in terms of $\al$ as follows
 \be
 &&|\la_{utr}\hb\equiv|\al_{utr}\hb+cycle(utr).
 \e{176}
$\al$ is only antisymmetric in the first two indices and have the index structure
\be
&&I^{(1)}_{rq}|\al_{tuv}\hb=-\del_{rt}|\al_{quv}\hb-\del_{ru}|\al_{tqv}\hb-\del_{rv}|\al_{tuq}\hb.
\e{1761}
The auxiliary field $\beta_{kqtr}$ is totally antisymmetric with the index structure
\be
&I_{ab}^{(1)}|\beta_{kqtr}\hb=\del_{a[k}|\beta_{qtr]b}\hb\equiv&\del_{ak}|\beta_{qtrb}\hb-\del_{aq}|\beta_{trkb}\hb+\nn\\&&+\del_{at}|\beta_{rkqb}\hb-\del_{ar}|\beta_{kqtb}\hb,
\e{1762}
which is the same  index structure as that of $|C_{kqtr}\hb$. Under gauge transformations it is required to transform as 
 \be
 &&\del|\beta_{uqtr}\hb={1\over4}\Big(d_ud_qd_t|\varepsilon_r\hb+cycle(uqtr)\Big).
 \e{177}
 (This cycle contains signs.)

  In the half-integer case the auxiliary field $\al_{kr}$ is required to satisfy the index structure
  \be
  &&I_{rq}^{(1)}|\al_{tu}\hb=-\del_{rt}|\al_{qu}\hb-\del_{ru}|\al_{tq}\hb.
  \e{1771}
  and to determine the compensator field $\la_{kr}$ in $|\cE_{\rho,\la}\hb$ through the relation
 \be
 &&|\la_{kr}\hb\equiv|\al_{kr}\hb-|\al_{rk}\hb.
 \e{179}
 Under gauge ransformations it is required to transform as 
 \be
 \del|\al_{kr}\hb=\tau_k|\xi_r\hb,
 \e{178}
which is consistent with the relation \rl{179}. 
 The auxiliary field $\beta_{tkr}$  is totally antisymmetric with the index structure
 \be
 &&I_{ab}^{(1)}|\beta_{tkr}\hb=-\del_{a[t}|\beta_{kr]b}\hb\equiv-\del_{at}|\beta_{krb}\hb-\del_{ak}|\beta_{rtb}\hb-\del_{ar}|\beta_{tkb}\hb,
 \e{1791}
  which is the same as the index structure of $|Z_{tkr}\hb$. Under gauge transformations $\beta_{tkr}$ is required to transform as
 \be
 &&\del|\beta_{tkr}\hb={1\over3}\Big(d_td_k|\xi_r\hb+cycle(tkr)\Big).
 \e{180}

 The gauge invariance of the actions in \rl{174} follow from the  transformation formulas \rl{175}, \rl{177}, \rl{178}, \rl{180} and the relations
 \be
 &&d_r\Big(1-{1\over4}T^{\dag}_{ut}T_{ut}\Big)=d_r-\half d^{\dag}_uT_{ur}-{1\over4}T^{\dag}_{ut}T_{ut}d_r,\nn\\
 &&d_r\Big(1-\tau^{\dag}_k\tau_k+{1\over3}\tau^{\dag}_k\tau_q^{\dag}(T_{qk}+\tau_k\tau_q)\Big)=d_r+\cp\tau_r-{1\over3}d^{\dag}_k\Big(T_{kr}+\tau_r\tau_k\Big)-\nn\\
 &&\qquad\qquad-\tau_k^{\dag}\Big({1\over3}\cp(\tau_r\tau_k+\tau_k\tau_r)+\tau_kd_r-{1\over3}\tau^{\dag}_q(T_{qk}+\tau_k\tau_q)d_r\Big),
 \e{181}
 together with the identities \rl{1611} and \rl{162}.
 
  Compared to the ingredients of the previously derived equations,  \rl{129} and \rl{137}, the $\al$ and $\beta$ fields are new objects although part of the $\al$ fields are related to the compensator fields $\la$. Below it will be shown that the new fields act like Lagrange multipliers in the actions \rl{174}.

  \subsection{Remarks on the Lagrangian equations of motion from \rl{174}}
 Since the actions $S_1$ and $S_2$ in \rl{174} are gauge invariant it follows that  their equations of motion are gauge invariant.

 A variation of $\beta_{uqtr}$ in $S_1$  and 
 a variation of $\beta_{tkr}$ in $S_2$ yield trivially
 \be
 &&|C_{uqtr}\hb=0,\nn\\
 &&|Z_{tkr}\hb=0,
 \e{190}
 respectively, where $C$ and $Z$ are defined in \rl{1621} and \rl{163}. These equations are gauge invariant. They were also argued for in subsection 4.1 and they yield the (Fang)-Fronsdal double trace conditions \rl{169} in the limit \rl{166}. 
 In order to analyse the equations that follow from a variation of $\al$ it is convenient to decompose $\al$ in terms of independent fields as follows
 \be
 &&|\al_{krt}\hb= {1\over3}\Big(|\la_{krt}\hb+|\ga_{krt}\hb-|\ga_{rkt}\hb\Big),\nn\\
 &&|\al_{rt}\hb=\half\Big(|\la_{rt}\hb+|\ga_{rt}\hb\Big),
 \e{1901}
 where the compensator fields $\la$ are given by \rl{176} and \rl{179}, and where
 \be
 &&|\ga_{krt}\hb\equiv |\al_{krt}\hb+|\al_{trk}\hb,\nn\\
 &&|\ga_{rt}\hb\equiv |\al_{rt}\hb+|\al_{tr}\hb.
 \e{1902}
 Since $|E_{\phi,\la}\hb$, $|\cE_{\rho,\la}\hb$, $|C_{uqtr}\hb$, and $|Z_{tkr}\hb$ are independent of these $\ga$ fields, it follows that $\ga$ acts as a Lagrange multiplier in the actions \rl{174}. It is easily seen  that a variation of the $\ga$'s yields the equations
 \be
 &&\Big(T_{ut}d_r+T_{rt}d_u\Big)|E_{\phi,\la}\hb=0,\nn\\
 &&\Big(d_r\tau_k+d_k\tau_r+{2\over3}\cp(\tau_r\tau_k+\tau_k\tau_r)\Big)|\cE_{\rho,\la}\hb-\nn\\
 &&-{1\over3}\tau^{\dag}_q\Big\{(T_{qk}+\tau_k\tau_q)dr+(T_{qr}+\tau_r\tau_q)d_k\Big\}|\cE_{\rho,\la}\hb=0,
 \e{1903}
 which are gauge invariant. The variations of the gauge fields $\phi$, $\rho$ and the compensator fields $\la$ are more involved. The straight-forward expressions are given in appendix C. To check whether or not  \rl{190}, \rl{1903} and the equations in appendix C  reduce to
  \be
 &&|E_{\phi,\la}\hb=0,\quad |\cE_{\rho,\la}\hb=0.
 \e{1904}
 is highly nontrivial. (This check is complicated even in the symmetric case \cite{Francia:2005bu}.) The reason is that in order to reduce the equations one has apart from using the algebra of operators in  \rl{112} also use the properties (valid for $k=1,\ldots,[s]$)
 \be
 &&(A_k)^{d/2}|\phi\hb=0,\quad (a_k)^{d/2}|\rho\hb=0,
 \e{1905}
 where the operators $A_k$ and $a_k$ represent any of the following operators 
 \be
 &&A_k=d_k, T_{jk}\;\;\;{\rm any}\; j,\quad a_k=d_k, \tau_k, T_{jk}\;\;\;{\rm any}\; j.
 \e{1906}
 Thus, in \eg $d=4$ (symmetric gauge fields) the product of any two operators given in \rl{1906} yields zero on the gauge states. Apart from these difficulties one should at least be able to rewrite the equations in appendix C in terms of a set of elementary gauge invariant states as is done in the symmetric case \cite{Francia:2005bu,Francia:2007qt}. In order to do so one has first to find a sufficient number of different elementary gauge invariant states. However, so far I have only given the following gauge invariant states: $|E_{\phi,\la}\hb$, $|\cE_{\rho,\la}\hb$, $|C_{uqtr}\hb$, $|Z_{tkr}\hb$ in \rl{1281}, \rl{1361}, \rl{1621} and  \rl{163} respectively. Guided by the similarity to the treatment of the symmetric case in \cite{Francia:2005bu,Francia:2007qt} I find also the following gauge invariant states
 \be
 &&|B_{uqtr}\hb\equiv|\beta_{uqtr}\hb-{1\over12}|K_{uqtr}\hb,
 \e{191}
 \be
 &&|B_{qrt}\hb\equiv|\beta_{qrt}\hb-{1\over6}|Y_{qrt}\hb,
 \e{1910}
 where
 \be
  &&|K_{uqtr}\hb\equiv d_{[u}d_qT_{tr]} |\phi\hb-d_k^{\dag}d_{[u}d_q|\al_{tr]k}\hb-p^2d_{[u}|\la_{qtr]}\hb,\nn\\
   &&|Y_{qrt}\hb\equiv d_{[q}T_{rt]}|\rho\hb-d^{\dag}_ud_{[q}\tau_r|\al_{t]u}\hb+p^2\tau_{[q}|\la_{rt]}\hb,
 \e{1911}
 where in turn the antisymmetric expressions may be written as
 \be
 &&d_{[u}d_qT_{tr]}\equiv d_ud_qT_{tr}-d_qd_tT_{ru}+cycle(qtr),\nn\\
 &&d_{[u}d_q|\al_{tr]k}\hb\equiv d_ud_q|\al_{trk}\hb-d_qd_t|\al_{ruk}\hb+cycle(qtr),\nn\\
 &&d_{[u}|\la_{qtr]}\hb\equiv d_u|\la_{qtr}\hb-d_q|\la_{tru}\hb+d_t|\la_{ruq}\hb-d_r|\la_{uqt}\hb,\nn\\
 &&d_{[q}T_{rt]}\equiv d_qT_{rt}+cycle(qrt),\nn\\
 &&d_{[q}\tau_r|\al_{t]u}\hb\equiv d_q(\tau_r|\al_{tu}\hb-\tau_t|\al_{ru}\hb)+cycle(qrt),\nn\\
 &&\tau_{[q}|\la_{rt]}\hb\equiv \tau_q|\la_{rt}\hb+cycle(qrt).
 \e{1912}
$|\la_{qtr}\hb$ and $|\la_{rt}\hb$ are given by \rl{176} and \rl{179}. The above expressions \rl{191} and \rl{1910}
may be compared to $\cal{B}$ and $\cal{Y}$ in the symmetric case given by (2.13) and (2.50) in \cite{Francia:2007qt}. In the symmetric case the equations of motion are possible to express in terms of the gauge invariant states $\cA, \cC, \cB$ (integer spins $s$) and $\cW, \cal{Z}, \cal{Y}$ (half-integer spins $s$). I have now what looks like the corresponding gauge invariant states (see table at the end of subsection 4.1 and above).
A further similarity to the symmetric case is that  the actions $S_1$ and $S_2$ involve the gauge invariant states 
\be
&&\Big(1-{1\over4}T^{\dag}_{ut}T_{ut}\Big)|E_{\phi,\la}\hb,\nn\\
&&\Big(1-\tau^{\dag}_k\tau_k+{1\over3}\tau^{\dag}_k\tau_q^{\dag}(T_{qk}+\tau_k\tau_q)\Big)|\cE_{\rho,\la}\hb,
\e{301}
which correspond to the  symmetric expressions  
\be
&&{\cal{A}}-{\half}\eta{\cal{A}'},\quad{\cal{W}}-{\half} \eta{\cal{W}'}-\half\ga{\cal{W}}\!\!\!\!\!\slash
\e{302}
in (2.7) and (2.47) in \cite{Francia:2007qt}. The gauge invariant states \rl{301} enter also the equations of motion for $\phi$ and $\rho$ in appendix C which is also the case for \rl{302} in the symmetric case.

Now the question is whether or not there are further elementary gauge invariant states here than those suggested by the symmetric case. One indication for this is the presence of the $\ga$-fields in the actions. 
There is nothing like the $\ga$-field in \cite{Francia:2005bu} and there are no equations like \rl{1903}. 
 The $\ga$-fields enter here already for $s\geq3/2$ in general. However,  in $d=4$ they enter at the same level as the compensator fields $\la$ due to the relations \rl{1905}. All this makes me believe that the Lagrangians contain redundant degrees of freedom that either have to be fixed by gauge fixing conditions or by  further gauge invariant conditions which then may be imposed by means of gauge invariant Lagrange multipliers, or they may be transformed away in some way.

If the Lagrangian equations can be written in a manifestly gauge invariant form using 
 only $|E_{\phi,\la}\hb$, $|\cE_{\rho,\la}\hb$, $|C_{uqtr}\hb$, $|Z_{tkr}\hb$ and the $|B\hb$-states \rl{191} and \rl{1910} as suggested by the corresponding analysis in the symmetric case in \cite{Francia:2005bu}, then  it follows that the  $\beta$-fields and the $\ga$-fields only enter  the gauge invariant states \rl{191} and \rl{1910}. In other words they  enter only in the combinations allowed by these expressions, \ie
 \be
 &&|\beta_{uqtr}\hb+{1\over36}d^{\dag}_kd_{[u}d_q\Big(|\ga_{trk}\hb-|\ga_{rtk}\hb\Big),\nn\\
 &&|\beta_{qrt}\hb+{1\over12}d^{\dag}_ud_{[q}\tau_r|\ga_{t]u}\hb.
 \e{304}
 This implies that the $\ga$-states are pure gauge states: transformations of the form $|\ga\hb\rightarrow|\ga\hb+|\theta\hb$ may be compensated by transformations of $|\beta\hb$. It follows then that one may transform away the $|\ga\hb$-states completely and replace the $|\beta\hb$-states by
 \be
 &&|\beta'_{uqtr}\hb=|\beta_{uqtr}\hb+{1\over36}d^{\dag}_kd_{[u}d_q\Big(|\ga_{trk}\hb-|\ga_{rtk}\hb\Big),\nn\\
 &&|\beta'_{qrt}\hb=|\beta_{qrt}\hb+{1\over12}d^{\dag}_ud_{[q}\tau_r|\ga_{t]u}\hb.
 \e{305}
 However, removing the $|\ga\hb$-states in the actions  and replacing the $|\beta\hb$-states by the above $|\beta'\hb$-states yields new actions which seems far from equivalent with the original actions which should be the case if the starting assumption is right. It follows then that there must exist further gauge invariant states than those constructed so far. In fact, examples of additional gauge invariant states are 
  \be
 &|H_{kqtru}\hb\equiv& T_{rk}|\la_{uqt}\hb+T_{rq}|\la_{utk}\hb+T_{rt}|\la_{ukq}\hb+T_{kq}|\la_{urt}\hb+T_{qt}|\la_{urk}\hb+\nn\\&&+T_{tk}|\la_{urq}\hb+T_{uk}|\la_{qtr}\hb+T_{qu}|\la_{trk}\hb+T_{ut}|\la_{rkq}\hb+T_{ru}|\la_{kqt}\hb-\nn\\&&-2T_{[kq}|\al_{tr]u}\hb,\nn\\
 &|H_{tkrq}\hb\equiv& T_{kr}|\la_{qt}\hb+T_{rt}|\la_{qk}\hb+T_{tk}|\la_{qr}\hb+\nn\\&&+\tau_q\Big(\tau_k|\la_{rt}\hb+\tau_r|\la_{tk}\hb+\tau_t|\la_{kr}\hb\Big)+T_{[tk}|\al_{r]q}\hb,
 \e{306}
where the first states are antisymmetric in $kqtr$ and the second in $tkr$. If $|H\hb=0$ is imposed as suggested by the following treatment then $|\la\hb$ and $|\ga\hb$ will be related contrary to the previous treatment.

In conclusion the  results here cast some doubts that the actions \rl{174} without modifications actually yield the required equations.  Since the reduction in terms of manifestly gauge invariant states is very difficult to do here, much more difficult than in the symmetric case in \cite{Francia:2005bu},  which is far from easy, I turn now to the somewhat simpler minimal formulation. (This formulation has several ingredients with similar structures as the ingredients in the extended symmetric treatment in \cite{Campoleoni:2008jq,Campoleoni:2009gs}. There are \eg corresponding relations to \rl{306}.)

  \setcounter{equation}{0}
\section{The minimal formulation}
The previous compensator fields appeared in the general solutions of the equations of motion, solutions which in turn are equations for the gauge fields. However, in order for the latter equations to be fully gauge invariant one does not need the most general form of the compensator fields that have been considered so far. There is in fact a much simpler minimal form that reproduce the same gauge transformations. To see this I rewrite the previous compensator fields as follows (cf (1.3) in \cite{Campoleoni:2008jq}, and (2.11) in \cite{Campoleoni:2009gs})
 \be
 &&|\la_{rqt}\hb=T_{rq}|X_t\hb+T_{qt}|X_r\hb+T_{tr}|X_q\hb,\nn\\
 &&|\la_{rq}\hb=\tau_r|\Xi_q\hb-\tau_q|\Xi_r\hb,
 \e{601}
 where $|X\hb$ and $|\Xi\hb$ are the minimal compensator fields which have an index structure   determined by the conditions
 \be
 &&I^{(1)}_{qr}|X_t\hb=-\del_{qt}|X_r\hb,\quad I^{(1)}_{qr}|\Xi_t\hb=-\del_{qt}|\Xi_r\hb.
 \e{60101}
 The required gauge transformations \rl{135} and \rl{141} follow if one demands that $|X\hb$ and $|\Xi\hb$  transform as
 \be
 &&\del|X_r\hb=|\varepsilon_r\hb,\nn\\
 &&\del|\Xi_r\hb=|\xi_r\hb,
 \e{602}
 where $|\varepsilon_r\hb$ and $|\xi_r\hb$ are the gauge parameters. The gauge transformations of $|X\hb$ and $|\Xi\hb$ are therefore trivial.
 Notice that the new auxiliary fields $\al$ introduced in the actions \rl{174} have the following simple form in terms of $|X\hb$ and $|\Xi\hb$,
\be
&&|\al_{utr}\hb=T_{ut}|X_r\hb,\nn\\
&&|\al_{kr}\hb=\tau_k|\Xi_r\hb.
\e{603}
In terms of $|X\hb$ and $|\Xi\hb$ one may now define new gauge invariant states \'a la St\"uckelberg: (cf (2.15) in \cite{Campoleoni:2008jq}, and (2.14) in \cite{Campoleoni:2009gs})
\be
 &&|\bphi\hb\equiv|\phi\hb-d^{\dag}_k|X_k\hb,\nn\\
&&|\brho\hb\equiv|\rho\hb-d^{\dag}_k|\Xi_k\hb,
\e{604}
which are consistent with the index structure \rl{122}, \ie
\be
&&I^{(1)}_{qr}|\bphi\hb=0,\quad  I^{(1)}_{qr}|\brho\hb=0.
\e{60401}
Indeed inserting \rl{601} into the $|E\hb$-states in the equations of motion, \ie  \rl{1281} and \rl{1361}, I find
\be
&&|E_{\phi,\la}\hb=|F_{\bphi}\hb\equiv\Big(p^2-d_r^{\dag}d_r-\half d_r^{\dag}d_t^{\dag}T_{rt}\Big)|\bphi\hb,\nn\\
&&|\cE_{\rho,\la}\hb=|\cF_{\brho}\hb\equiv\Big(\cp-d^{\dag}_r\tau_r\Big)|\brho\hb,
\e{605}
where $|F\hb$, $|\cF\hb$ are   the Fronsdal states \rl{167}. (Notice that $|F_{\bphi}\hb=|F_{\phi}\hb$ for $s=1,2$, and $|\cF_{\brho}\hb=|\cF_{\rho}\hb$ for $[s]=1$.)      These states are now trivially gauge invariant due to \rl{604}. All identities in section 4 are still valid when \rl{601} is inserted. Notice that $|C\hb$ and $|Z\hb$ in \rl{1621} and \rl{163} reduce to (cf (2.20) in \cite{Campoleoni:2008jq}, and (2.17) in \cite{Campoleoni:2009gs})
\be
&&|C_{kqtr}\hb=(TT)_{kqtr}|\bphi\hb,\nn\\
&&|Z_{krt}\hb=(T\tau)_{krt}|\brho\hb,
\e{606}
which are manifestly gauge invariant due to \rl{604}.

Also the $|\beta\hb$-fields in the actions \rl{174} have corresponding gauge invariant expressions like \rl{604}. They are
\be
&&|\bbeta_{uqtr}\hb\equiv|\beta_{uqtr}\hb-{1\over4}d_{[u}d_qd_t|X_{r]}\hb,\nn\\
&&|\bbeta_{qrt}\hb\equiv|\beta_{qrt}\hb-{1\over3}d_{[q}d_r|\Xi_{t]}\hb,
\e{607}
which are consistent with the index structures \rl{1762} and \rl{1791},\ie
\be
&&I_{ab}^{(1)}|\bbeta_{kqtr}\hb=\del_{a[k}|\bbeta_{qtr]b}\hb,\quad I_{ab}^{(1)}|\bbeta_{tkr}\hb=-\del_{a[t}|\bbeta_{kr]b}\hb.
\e{6071}
Notice that the gauge invariant states $|B\hb$ in \rl{191} and \rl{1910} reduce to the following expressions (these operators also satisfy \rl{6071})
\be
&&|B_{uqtr}\hb=|\bbeta_{uqtr}\hb-{1\over12}d_{[u}d_qT_{tr]}|\bphi\hb,\nn\\&& |B_{qrt}\hb=|\bbeta_{qrt}\hb-{1\over6}d_{[q}T_{rt]}|\brho\hb,
\e{608}
(The antisymmetric operator expressions are given explicitly in \rl{1912}.)
The gauge invariant states $|H\hb$ in \rl{306} reduce simply to zero when \rl{601} is inserted.

The actions $S_1$ and $S_2$ in \rl{174} become after insertion of \rl{601} the following manifestly gauge invariant actions
\be
&S_1=&\half\vb\bphi|\Big(1-{1\over4}T^{\dag}_{ut}T_{ut}\Big)|F_{\bphi}\hb+{1\over24}\vb\bbeta_{uqtr}|(TT)_{uqtr}|\bphi\hb+c.c.,\nn\\
&S_2=&\half\vb\bphi|\Big(1-\tau^{\dag}_k\tau_k+{1\over3}\tau^{\dag}_k\tau_q^{\dag}(T_{qk}+\tau_k\tau_q)\Big)|\cF_{\brho}\hb-\nn\\&&-{1\over12}\vb\bbeta_{qrt}|(T\tau)_{qrt}|\brho\hb+c.c..
\e{609}
The manifestly gauge invariant equations from these actions are easy to derive. From a variation of $\bbeta$ one finds
\be
&&(TT)_{uqtr}|\bphi\hb=0,\nn\\
&&(T\tau)_{qrt}|\brho\hb=0,
\e{610}
and a variation of $|\bphi\hb$ and $|\brho\hb$ yield
\be
&&\Big(1-{1\over4}T^{\dag}_{ut}T_{ut}\Big)|F_{\bphi}\hb+{1\over24}(T^{\dag}T^{\dag})_{uqtr}|\bbeta_{uqtr}\hb-R_1=0\nn\\
&&\Big(1-\tau^{\dag}_k\tau_k+{1\over3}\tau^{\dag}_k\tau_q^{\dag}(T_{qk}+\tau_k\tau_q)\Big)|\cF_{\brho}\hb-{1\over12}(T^{\dag}\tau^{\dag})_{qrt}|\bbeta_{qrt}\hb-\cR_2=0,\nn\\
\e{611}
where $R_1$ is given in \rl{632}, and $\cR_2$ in \rl{636} below.
However, even within this simple minimal formulation it is not easy to reduce these equations to a simple form.

\subsection{Proposals for  appropriate actions within the minimal formulation}
As an ansatz for the appropriate gauge invariant actions 
I propose the following generic forms 
\be
&\bar{S}_1=&\half\vb\bphi|Q|F_{\bphi}\hb+{1\over24}\vb B_{uqtr}|(TT)_{uqtr}|\bphi\hb+c.c.,\nn\\
&\bar{S}_2=&\half\vb\bphi|\cQ|\cF_{\brho}\hb-{1\over12}\vb B_{qrt}|(T\tau)_{qrt}|\brho\hb+c.c.,
\e{612}
where the gauge invariant states $|B\hb$ have the generic forms
\be
&&|B_{uqtr}\hb=|\bbeta_{uqtr}\hb+A_{uqtr}|\bphi\hb,\nn\\
&&|B_{tkr}\hb=|\bbeta_{tkr}\hb+A_{tkr}|\brho\hb,
\e{613}
where in turn the gauge invariant $|\bbeta\hb$-states are defined in \rl{607}. The $A$-operators are arbitrary apart from being totally antisymmetric and that they satisfy
\be
&&[I^{(1)}_{ab}, A_{uqtr}]=\del_{a[u}A_{qtr]b},\nn\\
&&[I^{(1)}_{ab}, A_{tkr}]=-\del_{a[t}A_{kr]b},
\e{61301}
which follows since the $|B\hb$-states \rl{613} must satisfy \rl{6071}.  Obviously $|B\hb=0$ for $s\leq3$. (The previous $|B\hb$-states \rl{608} are special examples of \rl{613}.)  The $Q$($\cQ$)-operators have the general form (cf the ans\"{a}tze for the projection operators \rl{203} in the following higher order theory)
\be
&&Q\equiv 1+\sum_{n=1}^M\sum_ib_{ni}(T^{\dag})^n_{(i)}(T)^n_{(i)},\nn\\
&&\cQ\equiv 1+\sum_{n=1}^M\sum_ic_{ni}(\tau^{\dag})^n_{(i)}(\tau)^n_{(i)},
\e{614}
where $T$ is the trace operator and $\tau$ the gamma-trace operator in \rl{111}. $b$ and $c$ are real parameters  determined by the conditions  below. Indices are summed over pairwise, one index on $T(\tau)$ and one index on $T^{\dag}(\tau^{\dag})$ and
for each order $n$ there are different ways to do the summation as indicated by the index $i$. 
For each finite value of the spin $s$ there is a finite integer $L$ such that
\be
&&(T)^{L}|F_{\bphi}\hb=0,\quad (\tau)^{L}|\cF_{\brho}\hb=0
\e{615}
due to the properties \rl{1905}, \rl{1906}.   The maximal order $M$ in \rl{614} is therefore finite  for each finite value of $s$.

Variations of $\bphi$ and $\brho$ in the actions \rl{612} yield the equations of motion 
\be
&&Q|F_{\bphi}\hb+(T^{\dag}T^{\dag})_{uqrt}|B'_{uqrt}\hb=R|\bphi\hb,\nn\\
&&\cQ|\cF_{\brho}\hb+(T^{\dag}\tau^{\dag})_{qrt}|B'_{qrt}\hb=\caR|\brho\hb,
\e{616}
where the $|B'\hb$-states also are of the form \rl{613} except that the $A$-operators might be different ($A'$). The operators $R$ and $\caR$ are given by  the expressions ($\Delta A\equiv A'-A$)
\be
&&R=\half[d^{\dag}_kd_k, Q]+{1\over4}\Big(T^{\dag}_{rq}d_qd_rQ+Qd^{\dag}_qd^{\dag}_rT_{rq}\Big)-{1\over24}(T^{\dag}T^{\dag})_{uqtr}\Delta A_{uqtr},\nn\\
&&\cR=-\half[\cp, \cQ]-\half\Big(\cQ d^{\dag}_r\tau_r-\tau^{\dag}_rd_r\cQ\Big)+{1\over12}(T^{\dag}\tau^{\dag})_{tkr}\Delta A_{tkr},
\e{61601}
which are nonzero  in general (see the explicit calculations below). In addition, the actions \rl{612} yield the equations \rl{610} from variations of $\bbeta$ for $s>3$. 

The constant parameters $b$ and $c$ in the ans\"atze for the $Q(\cQ)$-operators  are now proposed to be determined by the following conditions:\\
\hspace*{5mm}1) $Q(\cQ)$ must be hermitian.\\
\hspace*{5mm}2) $Q(\cQ)$ is required to satisfy the properties 
\be
&&[(TT)_{qrtk}, Q]|F_{\bphi}\hb=M_{qrtk;abcd}(TT)_{abcd}|F_{\bphi}\hb,\nn\\
&&[(T\tau)_{qrt}, \cQ]|\cF_{\brho}\hb=N_{qrt;abc}(T\tau)_{abc}|\cF_{\brho}\hb,
\e{617}
where $M$ and $N$ are operator matrices.\\
\hspace*{5mm}3) $Q(\cQ)$ should be chosen such that 
\be
&&R|\bphi\hb=0,\quad \caR|\brho\hb=0,
\e{618}
\ie the right-hand side in the equations \rl{616} should vanish. To obtain the equalities \rl{618} one should also use equations \rl{610} and the freedom to choose $|B'\hb$ (explicit in \rl{61601}).

If there are $Q(\cQ)$-operators satisfying the above conditions then I expect that the equations of motion \rl{616} from the actions \rl{612} yield the equations
\be
&&|F_{\bphi}\hb=0,\quad |\cF_{\brho}\hb=0,
\e{619}
and
\be
&&|B'_{uqtr}\hb=0,\quad |B'_{qrt}\hb=0.
\e{620}
The latter conditions determine the $\beta$-fields. The actual form of the $A'$-operators are  irrelevant. 

The reduction of the Lagrangian equations \rl{610} and \rl{616} to \rl{619} and \rl{620} is expected to proceed as follows: First one multiplies \rl{616} by $(TT)$ and $(T\tau)$ respectively. Then the identities \rl{1641} and \rl{16411} together with the properties 2) and 3) above make the equations \rl{616} reduce to
\be
&&(TT)_{abcd}(T^{\dag}T^{\dag})_{uqrt}|B'_{uqrt}\hb=0,\nn\\
&&(T\tau)_{abc}(T^{\dag}\tau^{\dag})_{qrt}|B'_{qrt}\hb=0.
\e{621}
These expressions may also be written as    
\be
&&|B'_{abcd}\hb+\al T^{\dag}_{kr}T_{[ab}|B_{cd]kr}\hb+\beta (T^{\dag}T^{\dag})_{uqrt}(TT)_{abcd} |B'_{uqrt}\hb=0,\nn\\
&&|B'_{abc}\hb+\al \tau^{\dag}_{k}\tau_{[a} |B'_{bc]k}\hb+\beta T^{\dag}_{kr}\tau_{[a}\tau_b|B'_{c]kr}\hb=0.
\e{622}
I expect then that by applying  $T$ and $\tau$ operators up to the limit allowed by \rl{1905},\rl{1906} it is possible to prove that \rl{621} reduce to \rl{620}. However, this remains to be proven. (The corresponding properties are proved in the symmetric case \cite{Francia:2005bu}.)  Anyway, if this is so it follows that the equations  \rl{616} reduce to
\be
&&Q|F_{\bphi}\hb=0,\quad\cQ|\cF_{\brho}\hb=0.
\e{623}
I expect then that there exist operators $Q'$ and $\cQ'$ of the same form as the ans\"{a}tze for $Q$ and $\cQ$ in \rl{614} which satisfy the properties
\be
&&Q'Q|F_{\bphi}\hb=|F_{\bphi}\hb,\quad\cQ'\cQ|\cF_{\brho}\hb=|\cF_{\brho}\hb.
\e{624}
By means of these $Q'(\cQ')$-operators the equations  \rl{623} then reduce to the wanted equations \rl{619}. For each finite $s$ and finite $M$ in \rl{614} there should be no problem to construct a $Q'(\cQ')$ of finite order satisfying \rl{624} due to the properties \rl{615} following from \rl{1905}, \rl{1906}.

Notice that exactly the same procedure applied to the actions \rl{612} in the (Fang)-Fronsdal limit $\bphi\ra\phi$, $\brho\ra\rho$ yields the generalized (Fang)-Fronsdal equations \rl{167}, \ie
\be
&&|F_{\phi}\hb=0,\quad |\cF_{\rho}\hb=0,
\e{625}
provided the procedure works. A partial indication that this is the case is given below.

 \subsection{Some explicit calculations}
 Here I shall consider some actions with definite explicit forms of the $Q(\cQ)$-operators in \rl{614}. Since it looks like the order $M$ in \rl{614} is determined by $N$ or equivalently the spin $s=N/2$ the expansions in the $Q(\cQ)$-operators in
\rl{614} should always be finite for a finite spin. I consider therefore the above formulations for  increasing orders $M$ in the $Q(\cQ)$-operators. (The order $M$ is denoted with an index in the following: $Q_M, \cQ_M$, $R_M$, $\cR_{M}$ and similarly for the corresponding actions: $(\bar{S}_1)_M, (\bar{S}_2)_M$.)
 
 To start with I consider the actions \rl{612} with the  zeroth order $Q(\cQ)$-operators, \ie for $Q_0=1$ and $\cQ_0=1$.  The actions \rl{612} become then
 \be
&(\bar{S}_1)_0=&\half\vb\bphi|F_{\bphi}\hb+{1\over24}\vb B_{uqtr}|(TT)_{uqtr}|\bphi\hb+c.c.,\nn\\
&(\bar{S}_2)_0=&\half\vb\bphi|\cF_{\brho}\hb-{1\over12}\vb B_{qrt}|(T\tau)_{qrt}|\brho\hb+c.c..
\e{626}
The equations of motion are \rl{610} and
\be
&&|F_{\bphi}\hb+{1\over24}(T^{\dag}T^{\dag})_{uqtr}| B_{uqtr}\hb=R|\bphi\hb,\nn\\
&&|\cF_{\brho}\hb-{1\over12}(T^{\dag}\tau^{\dag})_{qtr}| B_{qtr}\hb=\cR|\brho\hb,
\e{627}
where
\be
R_0={1\over4}\Big(T^{\dag}_{rq}d_qd_r+d^{\dag}_rd^{\dag}_qT_{qr}\Big),\quad\cR_0=\half\Big(\tau^{\dag}_rd_r-d^{\dag}_r\tau_r\Big).
\e{628}
 The conditions \rl{618} are only satisfied for $s\leq1$ in $d=4$. Indeed, for $s=1$ in $d=4$ \rl{626} reduces to
 \be
 &&(\bar{S}_1)_0=\half\vb\bphi|F_{\bphi}\hb+\half\vb F_{\bphi}|\bphi\hb=\vb F|F\hb,
 \e{6281}
 which is the conventional spin one action ($F$ is the original field strength according to \rl{120}).
 Notice that  $|F_{\bphi}\hb= |F_{\phi}\hb$ and $d_1|F_{\phi}\hb=0$. (There are no $|B\hb$-states here.)

Consider then the action \rl{612} for the first order $Q(\cQ)$-operators
\be
&&Q_1=1-{1\over4}T^{\dag}_{ut}T_{ut},\quad \cQ_1=1-\tau_r^{\dag}\tau_r.
\e{629}
These expressions satisfy conditions 1), hermiticity and 2), \rl{617}. The actions \rl{612} become $S_1$ in \rl{609} and
\be
&(\bar{S}_2)_1=&\half\vb\bphi|\Big(1-\tau^{\dag}_k\tau_k\Big)|\cF_{\brho}\hb-{1\over12}\vb\bbeta_{qrt}|(T\tau)_{qrt}|\brho\hb+c.c..
\e{630}
The equations of motion are \rl{610} and 
\be
&&Q_1|F_{\bphi}\hb+{1\over24}(T^{\dag}T^{\dag})_{uqtr}| B_{uqtr}\hb=R_1|\bphi\hb,\nn\\
&&\cQ_1|\cF_{\brho}\hb-{1\over12}(T^{\dag}\tau^{\dag})_{qtr}| B_{qtr}\hb=\cR_1|\brho\hb,
\e{631}
where
\be
&&R_1=-{1\over16}\Big(T^{\dag}_{ut}T^{\dag}_{rq}d_qd_rT_{ut}+d^{\dag}_rd^{\dag}_qT^{\dag}_{ut}T_{ut}T_{qr}\Big),\nn\\&&\cR_1=\half\Big(\tau^{\dag}_rd^{\dag}_r\tau_k\tau_r-\tau^{\dag}_r\tau^{\dag}_kd_r\tau_k\Big).
\e{632}
Condition 3),  \rl{618}, is satisfied for for $s\leq2$ in $d=4$, and for $s\leq1$ in  $d=6$, for $R_1$. For $\cR_1$ it is  satisfied for $s=1/2$ in arbitrary even dimensions $d$, and for $s\leq3/2$ in $d=4$. ($|B\hb=0$ for these $s$ values.)

Consider then the action \rl{612} for the second order $Q(\cQ)$-operators
\be
&&Q_2=1-{1\over4}T^{\dag}_{ut}T_{ut}-{1\over24}T_{ab}^{\dag}T_{cd}^{\dag}T_{ad}T_{cb},\nn\\
&& \cQ_2=1-\tau_r^{\dag}\tau_r+{1\over3}\tau^{\dag}_k\tau_q^{\dag}(T_{qk}+\tau_k\tau_q).
\e{633}
These expressions satisfy conditions 1), hermiticity, and 2), \rl{617}. The actions \rl{612} become  
\be
&(\bar{S}_1)_2=&\half\vb\bphi|Q_2|F_{\bphi}\hb+{1\over24}\vb B_{uqrt}|(TT)_{uqrt}|\bphi\hb+c.c.,
\e{634}
and $S_2$ in \rl{609}.
The equations of motion are \rl{610} and
\be
&&Q_2|F_{\bphi}\hb+{1\over24}(T^{\dag}T^{\dag})_{uqtr}| B'_{uqtr}\hb=R_2|\bphi\hb,\nn\\
&&\cQ_2|\cF_{\brho}\hb-{1\over12}(T^{\dag}\tau^{\dag})_{qtr}| B_{qtr}\hb=\cR_2|\brho\hb,
\e{635}
where
\be
&&R_2={1\over192}T^{\dag}_{ab}T^{\dag}_{cd}\Big(T^{\dag}_{rq}d_qd_r+d^{\dag}_qd^{\dag}_rT_{rq}\Big)T_{ab}T_{cd},\nn\\
&&\cR_2={1\over6}\tau^{\dag}_r\tau^{\dag}_k\tau^{\dag}_qd_r\Big(2\tau_k\tau_q-\tau_q\tau_k\Big)-{1\over6}\Big(2\tau^{\dag}_k\tau^{\dag}_q-\tau^{\dag}_q\tau^{\dag}_k \Big)   d^{\dag}_r\tau_k\tau_q\tau_r.\nn\\
\e{636}
This value of $R_2$ is obtained for the following value of $A'$ (cf \rl{608}):
\be
&&A'_{abcd}=A_{abcd}-{1\over12}d_{[a}d_bT_{cd]},
\e{637}
where the right-hand side is explicitly given in \rl{1912}. ($R_2$ also contains terms of the type $C_{abcd}(TT)_{abcd}$ which vanish in \rl{635} due to \rl{610}.)
Here condition 3), \rl{618}, is satisfied for spins $s\leq5$ in $d=4$, and $s\leq3$ in $d=6$ ($R_2$), and for  $s\leq3/2$ in $d=4,6$, and $s\leq5/2$ in $d=4$ ($\cR_2$).

Condition 3) \rl{618}, is obviously satisfied for higher and higher spins when higher and higher order $Q(\cQ)$-operators are used. Whether or not it will end for a finite order $M$ is unclear.

 \setcounter{equation}{0}
\section{BRST quantization}
The quantization chosen here is a straight-forward Dirac quantization. However, this quantization agrees completely with the BRST quantization in \cite{Marnelius:1988ab}, which was the starting point. The reason for this equivalence is that the BRST quantization in \cite{Marnelius:1988ab}  was performed on bilinear forms. 

A proper BRST quantization should be performed on an inner product space. In fact, the path integral formulation requires such a formulation. The path integrals may on the other hand always be expressed within the operator formulation which  provides for some deeper insights. First, a BRST quantization on an inner product space requires the number of unphysical variables to contain a factor four: there must be equally many odd as even unphysical variables, and half of each must be quantized with indefinite metric states. This is the generic case. In order to make sure that the number of unphysical variables contain a factor four one needs the extended BRST-BFV charge which contains dynamical Lagrange multipliers and antighosts. (Only the minimal BRST-BFV charge was used in  \cite{Marnelius:1988ab}.) 

In BRST quantizations one usually considers canonical theories in which one requires that every constraint may be gauge fixed classically. (This concept was introduced in \cite{Hwang:1988ak}.) However, the $O(N)$-extended supersymmetric particle model contain the $O(N)$ constraints $\psi_k\cdot\psi_l$ in \rl{102} which are of a non-canonical type. There are no gauge fixing variables to $\psi_k\cdot\psi_l$ which in turn is connected to the fact that there are no Newton-Wigner position operators \cite{Newton:1949cq} for massless particles with spin $s\geq1$. In the quantum theory here these non-canonical constraints are given by (4), (5), (6) in \rl{113} (traces and index structures), and (8), (9) in \rl{114} (gamma traces). 

In the quantum theory there is no problem to impose the non-canonical conditions in a BRST-invariant form without affecting the inner product properties. This operator formulation for path integral for particles with spins $s=1,1/2,1$ were investigated in \cite{Marnelius:1993ba}. The above rules for unphysical variables were shown to be valid, but it leaves some freedom. There are basically two options:  one may either choose time to be represented by a positive metric state space in which case the results should be very much the same as here. However, there is also the option to let time be represented by an indefinite metric state space. In this case the wave functions are defined on a Euclidean space and a BRST quantization leads to propagators (see \cite{Marnelius:1993ba} for details).  The spin one case is non-canonical and shown to require the treatment above. In fact, the conventional path integral treatment of the $O(N)$-extended supersymmetric particle model given in \cite{Pierri:1990rp} for arbitrary spins are in agreement with the above results. (In \cite{Pierri:1990rp} propagators are derived.)
 
 Now one should be able to perform a BRST quantization
 \be
 &&Q|\cF\hb=0,
 \e{192}
 where $Q$ is the extended BRST-BFV charge operator and where $|\cF\hb$ is a general state belong to an inner product space which depends on ghosts, antighosts and Lagrange multipliers. In this case time must be chosen with real eigenvalues in order for \rl{192} to have solutions. The physical states are contained in the states with ghost number zero. I expect the solutions of \rl{192} to be in agreement with the Dirac quantization here. The equation \rl{192} should be possible to imbed in an action principle by means of an action of the form
 \be
 &&S=\sum_n\vb\cF_n|Q|\cF_{-n-1}\hb,
 \e{193}
 where $n$ is the ghost number. ($Q$ has ghost number one and $S$ must have ghost number zero.) Alternatively one could have
  \be
 &&S=\vb\cF_{-1/2}|Q|\cF_{-1/2}\hb,
 \e{194}
 as in string field theory if $Q$ is in a minimal form or a gauge fixed form. Notice that these actions are of second order. To be precise one should insert an index $s$ on $Q$ and $\cF$ to indicate that there is a specific BRST charge $Q$ and an ansatz $|\cF\hb$ for each spin $s$. A more general action is then
 \be
 &&S=\sum_s\vb\cF_{s}|Q|\cF_{s}\hb,
 \e{195}
 where the sum may be chosen over any set of  different spins. (Ghost numbers may be chosen according to \rl{193} or \rl{194}.)  Eq.\rl{192} and the above actions are just suggestions. In order to find out whether these possibilities could be realized one has  to do a detailed analysis. Anyway all actions \rl{193}-\rl{195} are gauge invariant under the transformations
 \be
 &&|\cF\hb\quad\longrightarrow\quad|\cF\hb+Q|\chi\hb,
 \e{196}
where  $|\chi\hb$ is an arbitrary state with ghost number one less than $|\cF\hb$. 

Now the gauge transformations \rl{196} are not the gauge transformations treated before but
essentially the $O(N)$-extended supersymmetry transformations. The reason is that the gauge fields are hidden in the above BRST quantization. Therefore the equation \rl{192} probably only describes the gauge invariant field strengths, and the path integrals in \cite{Pierri:1990rp,Marnelius:1993ba} are probably only propagators for the field strengths (explicitly stated in \cite{Pierri:1990rp}). This BRST approach is therefore perhaps not what one is looking for here.

BRST quantization has otherwise been used in the higher spin theory for a very long time. However, it has then mainly been string inspired treatments using bosonic oscillators. (See \eg the recent reviews \cite{Buchbinder:2005cf,Fotopoulos:2008ka}.)

 \setcounter{equation}{0}
\section{Further remarks on the exact quantum theory}
I have derived free higher spin field theories from the $O(N)$ extended supersymmetric particle model. One may wonder if it is possible to do a similar derivation from another spinning particle model leading to an irreducible representation of the Poincar\'e group.  This cannot be excluded. (Bosonic oscillator models have \eg been the dominant ingredients in higher spin field theory.) However, in my opinion the $O(N)$ model is by far the most natural geometrical model on the market. There seems to exist no simple alternative model \cite{Marnelius:1990de}. On the other hand, it is doubtful whether the exact theory here can be used to find interacting higher spin theories. Probably one has a better chance if one then  starts from a multiplet model. Strings are, of course,   very geometrical multiplet models for which  interactions have been demonstrated to exist.

Still further remarks are given in section 13.

\part{The higher order quantum theory}
 \setcounter{equation}{0}
\section{Outline of the main results for the higher order quantum  theory}
The higher order theory defined in \cite{Marnelius:2008er} is a conformal theory in which the gauge fields for $s\geq2$ satisfy equations of higher order than two. In $d=4$ this theory was also to a large degree shown to fit into the framework proposed by Fradkin,Tseytlin and Linetski \cite{Fradkin:1985am,Fradkin:1989md,Fradkin:1990ps}, a relation which is further strengthened here.  (The index structure is different but equivalent \cite{Marnelius:2008er}.) Higher order equations are of course bad but the main motivation to consider these conformal theories is  due to the fact that they seem to allow for interactions: Such fields are \eg allowed as external fields in $d=4$ \cite{Arvidsson:2006}, and Fradkin and Linetski have even constructed cubic interactions  in a Lagrangian form in $d=4$ \cite{Fradkin:1989md,Fradkin:1990ps}.  The higher order theory in \cite{Marnelius:2008er} was extracted from the quantum $O(N)$ extended supersymmetric particle model in \cite{Marnelius:1988ab}, \ie exactly the model considered in the first part here. 
The starting point is to define a field with the same index structure as before, \ie the index condition (6) in \rl{113} is retained:
\be
&&I_{rq}|F\hb=0,\quad I_{rq}|\Psi\hb=0,\quad r,q=1,\ldots,[s].
\e{200}
The ans\"atze for $|F\hb$ and $|\Psi\hb$ are still given by \rl{106} and \rl{108} respectively. Gauge fields may be introduced by solving condition (3) in \rl{113}
\be
&&d^{\dag}_r|F\hb=0,\quad d^{\dag}_r|\Psi\hb=0,\quad r=1,\ldots,[s],
\e{201}
or (2) in \rl{113}
\be
&&d_r|F\hb=0,\quad d_r|\Psi\hb=0,\quad r=1,\ldots,[s],
\e{2011}
as in subsections 3.1 and 3.2. The  conditions in \rl{201} are solved by \rl{120}, and \rl{2011} is solved by \rl{1271}. As before these two solutions are dual to each other according to the rules in appendix B. The conditions \rl{200} are then solved by  \rl{122} and  \rl{1273} respectively.

 By means of the states $|F\hb$ and $|\Psi\hb$  I define the Weyl states $|C\hb$ and $|\cC\hb$  by
\be
&&|C\hb\equiv P|F\hb,\quad|\cC\hb\equiv \cP|\Psi\hb,
\e{202}
where the hermitian projections operators $P$ and $\cP$ have the structure
\be
&&P\equiv 1+\sum_{n=1}^M\sum_i\al_{ni}(T^{\dag})^n_{(i)}(T)^n_{(i)},\nn\\
&&\cP\equiv 1+\sum_{n=1}^M\sum_i\beta_{ni}(\tau^{\dag})^n_{(i)}(\tau)^n_{(i)},
\e{203}
where $T$ is the trace operator and $\tau$ the gamma-trace operator in \rl{111}. $\al$ and $\beta$ are real parameters  determined by the conditions \rl{205} below. For each order $n$ there are different ways to do the summation as indicated by the index $i$. The maximal order $M$ is finite since 
for each finite value of $s$ there is a finite integer $L$ such that
\be
&&(T)^{L}|F\hb=0,\quad (\tau)^{L}|\Psi\hb=0.
\e{204}
The parameters $\al$ and $\beta$ in \rl{203} are  determined by the conditions
\be
&&T_{rq}|C\hb=0,\quad \tau_r|\cC\hb=0,\quad r,q=1,\ldots,[s],
\e{205}
which implies that the corresponding Weyl tensors to $|C\hb$ and $|\cC\hb$ are traceless and gamma traceless respectively. 
The Weyl states \rl{202} are also required to be of such a form as to satisfy the index condition (6) in \rl{113}, \ie
\be
&&I_{rq}|C\hb=0,\quad I_{rq}|\cC\hb=0,\quad r,q=1,\ldots,[s].
\e{206}
It is suggested that the forms  \rl{202} satisfying \rl{205} and \rl{206} also satisfy the conditions
\be
&&T^{\dag}_{rq}|C\hb=0,\quad \tau^{\dag}_r|\cC\hb=0,\quad r,q=1,\ldots,[s].
\e{207}
The Weyl states satisfy therefore three  of the six conditions in \rl{113}, namely (4)-(6), and for half-integer spins also two  of the three conditions in \rl{114}, namely (8),(9). 
The duals of the Weyl states satisfy the same conditions. Furthermore, the Weyl states satisfy a generalized Weyl invariance which is an additional gauge invariance. Some details are given in sections 9 and 10 below. Conformally invariant actions are essentially the scalar products of the Weyl states. These actions which define the higher order theory are given in section 11.

 \setcounter{equation}{0}
\section{The Weyl states for integer spins}
\subsection{The projection operator $P$ for the Weyl states $|C\hb$ in \rl{202}}
For integer spins I start with the field $F$ with the index structure that follows from \rl{200}. Then I introduce the Weyl tensor $C$ by \rl{202}. The precise form of the projection operator $P$ in \rl{203}  is then up to fourth order ($P=1$ for $s\leq1$.)
\be
&&P=1+\al_1\sum_{r_1r_2}T_{r_1r_2}^{\dag}T_{r_1r_2}+\al_{21}\sum_{r_1r_2r_3r_4}T_{r_1r_2}^{\dag}T_{r_3r_4}^{\dag}T_{r_3r_4}T_{r_1r_2}+\nn\\
&&+\al_{22}\sum_{r_1r_2r_3r_4}T_{r_1r_2}^{\dag}T_{r_3r_4}^{\dag}T_{r_1r_4}T_{r_2r_3}+\nn\\&&+\al_{31}\sum_{r_1\ldots r_6}T_{r_1r_2}^{\dag}T_{r_3r_4}^{\dag}T_{r_5r_6}^{\dag}T_{r_5r_6}T_{r_3r_4}T_{r_1r_2}+\nn\\
&&+\al_{32}\sum_{r_1\ldots r_6}T_{r_1r_2}^{\dag}T_{r_3r_4}^{\dag}T_{r_5r_6}^{\dag}T_{r_5r_6}T_{r_2r_3}T_{r_4r_1}
+\nn\\&&+\al_{33}\sum_{r_1\ldots r_6}T_{r_1r_2}^{\dag}T_{r_3r_4}^{\dag}T_{r_5r_6}^{\dag}T_{r_6r_1}T_{r_4r_5}T_{r_2r_3}+\nn\\
&&+\al_{41}\sum_{r_1\ldots r_8}T_{r_1r_2}^{\dag}T_{r_3r_4}^{\dag}T_{r_5r_6}^{\dag}T_{r_7r_8}^{\dag}T_{r_7r_8}T_{r_5r_6}T_{r_3r_4}T_{r_1r_2}+\nn\\
&&+\al_{42}\sum_{r_1\ldots r_8}T_{r_1r_2}^{\dag}T_{r_3r_4}^{\dag}T_{r_5r_6}^{\dag}T_{r_7r_8}^{\dag}T_{r_8r_5}T_{r_6r_7}T_{r_4r_1}T_{r_2r_3}+\nn\\
&&+\al_{43}\sum_{r_1\ldots r_8}T_{r_1r_2}^{\dag}T_{r_3r_4}^{\dag}T_{r_5r_6}^{\dag}T_{r_7r_8}^{\dag}T_{r_7r_8}T_{r_5r_6}T_{r_4r_1}T_{r_2r_3}+\nn\\
&&+\al_{44}\sum_{r_1\ldots r_8}T_{r_1r_2}^{\dag}T_{r_3r_4}^{\dag}T_{r_5r_6}^{\dag}T_{r_7r_8}^{\dag}T_{r_7r_8}T_{r_6r_1}T_{r_2r_3}T_{r_4r_5}+\nn\\
&&+\al_{45}\sum_{r_1\ldots r_8}T_{r_1r_2}^{\dag}T_{r_3r_4}^{\dag}T_{r_5r_6}^{\dag}T_{r_7r_8}^{\dag}T_{r_8r_1}T_{r_2r_3}T_{r_4r_5}T_{r_6r_7}+O((T^{\dag})^5T^5).\nn\\
\e{210}
The terms in this expansion is determined by the conditions $P=P^{\dag}$ and that all indices are summed over in the following fashion: Each index appears in pairs: one index is on a $T^{\dag}$ and one on a $T$. The maximal order $N$ is determined by the property
\be
&&(T_{a\cdot})^{d/2+1}|F\hb=0
\e{211}
for any fixed index value $a$ (the dot denotes any other index value which may be different for each factor in the product)

I believe there are three ways to determine the $\al$-parameters in $P$: 1) $C$ is traceless, 2) $C$ is invariant under generalized Weyl transformations, 3) $P$ is a projection operator. I start with the first way
  \beq
 \begin{tabular}{|p{114mm}|}
 \hline
  {\bf Proposal 1}: The real parameters $\al_{ki}$ in \rl{210} are determined by the conditions
 \be
 &&T_{qr}|C\hb=0,\quad q,r=1,\ldots,s.
 \e{215}\\
  \hline
 \end{tabular}
 \eq
 Note that this property implies that the Weyl tensor $C$ is traceless.

 By means of the algebra \rl{112} I find to lowest orders
 \be
 &&T_{qr}|C\hb=\Big(1+4\al_1(s-1)\Big)T_{qr}|F\hb+\nn\\&&+\Big(\al_1+8\al_{21}(s-1)-4\al_{22}\Big)\Big(\sum_{t,u=1}^sT^{\dag}_{tu}T_{tu}\Big)T_{qr}|F\hb-\nn\\&&-
 \Big(4\al_{21}-2(2s-3)\al_{22}\Big)\sum_{t,u=1}^sT^{\dag}_{tu}(T_{qt}T_{ur}-T_{qu}T_{tr})|F\hb+\nn\\&&+O\Big((T^{\dag})^2T^3\Big)|F\hb.
 \e{216}
 Hence, conditions \rl{215} yields
 \be
 &&\al_1=-{1\over4(s-1)},\nn\\
 &&\al_{21}={2s-3\over32(s-1)C_s},\nn\\
 &&\al_{22}={1\over16(s-1)C_s},\nn\\
 &&C_s\equiv(2s-3)(s-1)-1.
 \e{217}
 These results are valid in {\em any} even dimension for $s>2$. However, \rl{217} is not valid for $s=2$, since $C_2=0$ then. The reason for this is that for $s=2$ the two second order terms in $P$ are equal. Notice that there is only one trace operator $T$ for $s=2$ ($T_{12}$).  I find here
 \be
&&T_{12}|C\hb=\Big(1+4\al_1\Big)T_{12}|F\hb+\nn\\&&+2\Big(\al_1+6\al_2\Big)T^{\dag}_{12}(T_{12})^2|F\hb+O\Big((T^{\dag}_{12})^2(T_{12})^3\Big)|F\hb,
 \e{218}
 where
 \be
 &&\al_2\equiv 2\al_{21}-\al_{22}.
 \e{219}
 For $s=2$ the condition \rl{215} yields therefore
 \be
 &&\al_1=-{1\over4},\quad \al_2={1\over24}
 \e{220}
 up to order two. For $d=4$ this is the maximal order due to \rl{211}.
 
 In [1] the Weyl tensors were explicitly  calculated up to $s=4$ in $d=4$. The ansatz there should be equivalent to the ansatz \rl{210} here.  For $s=2,3$ they seem to be the same. Notice that the maximal order is two even for $s=3$ due to the equality
 \be
 &&T_{12}T_{13}T_{23}|F\hb=0,
 \e{221}
 which follows since the total trace for odd spins always yields zero. (This expression is a total trace of $F$ since all indices appear twice.) This implies that the maximal order is $s-1$ for odd $s$, and $s$ for even $s$ in $d=4$. Now for even $s$ not all terms are different  in the maximum order as was shown for $s=2$ above. For $s=4$ in $d=4$  there are formally 11 parameters in $P$.  However, like for $s=2$ it is natural to expect that  not all terms are different  in the maximum orders. Comparisons with the explicit expression in \cite{Marnelius:2008er} reveal that the relation to $P$ above is even more complex than that. In fact, it seems as if $P$ in \rl{210}  makes two of the parameters in the second order in \cite{Marnelius:2008er} dependent. Furthermore,  only two parameters in $P$ in the 4th order are independent and only two of the three parameters in the third order. Thus, for $s=4$ in $d=4$ $P$ contains only 7 independent parameters to be determined by \rl{215}. For $s=5$ in $d=4$ there are at most 11 independent parameters from the expression \rl{210} since the fourth order is the maximal order even here.

 \subsection{The generalized Weyl invariance}
 \beq
 \begin{tabular}{|p{114mm}|}
 \hline
  {\bf Proposal 2}: The Weyl state satisfying \rl{215} is also invariant under the generalized Weyl transformation
    \be
  &&|F\hb\;\;\longrightarrow\;\;|F\hb+\sum_{q,r=1}^sT^{\dag}_{qr}|\zeta_{qr}\hb,
  \e{222}
  where the antisymmetric state $|\zeta_{qr}\hb$ has the index structure
  \be
  &&I_{qr}|\zeta_{tu}\hb=\del_{qt}|\zeta_{ur}\hb-\del_{qu}|\zeta_{tr}\hb.
  \e{223}
  Vice versa if $|C\hb$ is invariant under \rl{222} then $|C\hb$ satisfies \rl{215}.\\
 \hline
 \end{tabular}
 \eq
 I find up to second order (cf \rl{216})
 \be
 &&\sum_{r,q=1}^sPT_{rq}^{\dag}|\zeta_{rq}\hb=\Big(1+4\al_1(s-1)\Big)\sum_{r,q=1}^sT_{rq}^{\dag}|\zeta_{rq}\hb+\nn\\
 &&+\Big(\al_1+8\al_{21}(s-1)-4\al_{22}\Big)\Big(\sum_{t,u=1}^sT_{qr}^{\dag}T^{\dag}_{tu}T_{tu}\Big)|\zeta_{rq}\hb-\nn\\
 &&-\Big(4\al_{21}-2(2s-3)\al_{22}\Big)\sum_{t,u=1}^s(T^{\dag}_{qt}T^{\dag}_{ur}-T^{\dag}_{qu}T^{\dag}_{tr})T_{tu}|\zeta_{rq}\hb+\nn\\
 &&+O\Big((T^{\dag})^3T^2|\zeta\hb\Big),
 \e{2231}
 which vanishes for the values \rl{217}.

 \subsection{The index structure}
  \beq
 \begin{tabular}{|p{114mm}|}
 \hline
  {\bf Theorem 1}: The projection operator $P$ satisfies the property
  \be
  &&[I_{qr}, P]=0,\quad  q,r=1,\ldots,s.
  \e{212}\\
 \hline
 \end{tabular}
 \eq
 
 Proof:  Consider the transformation of one index pair (t)
 \be
 &&[I_{qr}, \sum_t\cdots T^{\dag}_{tu}\cdots T_{tv}\cdots]=\nn\\&&\sum_t\Big\{-\del_{rt}(\cdots T^{\dag}_{uq}\cdots T_{rv}\cdots)+\del_{qt}(\cdots T^{\dag}_{uq}\cdots T_{rv}\cdots)\Big\}+\cdots\nn\\&&=0+\cdots,
 \e{213}
 where the last dots represent terms coming from the transformation of all the other index pairs apart from $t$. Since one pair yields zero all pairs yield zero $\bullet$
 
 It follows that
 \be
 &&I_{qr}|C\hb=0,
 \e{214}
 since $I_{qr}|F\hb=0$. Thus, the Weyl tensor $C$ has the same index structure as $F$ for arbitrary even $d$. (For d=4 this was given as a proposal in \cite{}.)

\subsection{$P$ as a projection operator}
From the form of $P$ in \rl{210} and the property \rl{215} it follows that 
\be
&&P|C\hb=|C\hb.
\e{2141}
$P$ should therefore satisfy the projection property
\be
&&P^2=P+\sum_{r,q=1}^s\Big(\quad\Big)_{rq}I_{rq}.
\e{2142}
However,  this relation when applied to the ansatz \rl{210} is not useful to determine the $\al$'s, since \rl{2142}  is nonlinear in the $\al$-parameters. (One solution of  \rl{2142} is of course also $P=1$.)

 \subsection{The dual projector $\tilde{P}$}
 The dual to $P$, $\tilde{P}$,  is according to appendix B equal to $P$ with the replacement $T \leftrightarrow -T^{\dag}$, which implies that $\tilde{P}$ has the structure (cf \rl{203})
 \be
&&\tilde{P}\equiv 1+\sum_{n=1}^N\sum_i\al_{ni}(T)^n_{(i)}(T^{\dag})^n_{(i)}.
\e{226}
  \beq
 \begin{tabular}{|p{114mm}|}
 \hline
  {\bf Proposal 3}: The Weyl state $|C\hb$ in \rl{202} may also be written as
      \be
  &&|C\hb=\tilde{P}|F\hb.
    \e{227}\\
 \hline
 \end{tabular}
 \eq
 This property requires that
 \be
 \sum[(T^{\dag})^n, T^n]=\sum_{q,r}\big(\cdots\big)_{qr}I_{qr}
 \e{228}
for each term in $P$ ($\tilde{P}$), which I have checked up to second order.

$\tilde{P}$ should then also be a projector
\be
&&\tilde{P}|C\hb=|C\hb.
\e{229}

A consequence of the relation \rl{227} is that the $\al$-parameters are also determined by the conditions (cf \rl{215} and \rl{2231})
 \be
 &&T^{\dag}_{qr}|C\hb=0,\quad q,r=1,\ldots,s,\nn\\
  &&\sum_{r,q=1}^sPT_{rq}|\tilde{\zeta}_{rq}\hb=0,
  \e{230}
  where $\tilde{\zeta}$ is the dual of $\zeta$. The index structure is
  \be
  &&I_{qr}|\tilde{\zeta}_{tu}\hb=\del_{rt}|\tilde{\zeta}_{qu}\hb-\del_{ru}|\tilde{\zeta}_{qt}\hb.
  \e{2301}
  The last equation in \rl{230} is the dual generalized Weyl invariance of $|C\hb$ under the transformations
  \be
  &&|F\hb\;\;\longrightarrow\;\;|F\hb+\sum_{q,r=1}^sT_{qr}|\tilde{\zeta}_{qr}\hb.
  \e{2302}
  
  From the relation \rl{227} it follows that the dual of $|C\hb$ satisfies the same properties as $|C\hb$. Thus, as on $|F\hb$ one may impose (anti)self-duality conditions on $|C\hb$ (if $F$ is (anti)self-dual $C$ is (anti)self-dual).

\subsection{The Weyl states in terms of gauge fields}
So far the Weyl states $|C\hb$ are expressed in terms of the field strengths $|F\hb$: $|C\hb=P|F\hb$. However, in \cite{Marnelius:2008er} the Weyl tensors were expressed in terms of gauge fields. Using the results of part I I define the gauge states $|\phi\hb$ through the relation \rl{120} \ie
\be
&&|F\hb\equiv \Pi|\phi\hb, \quad \Pi\equiv  d^{\dag}_1d^{\dag}_2\cdots d^{\dag}_{s}.
\e{231}
This implies then
\be
&&|C\hb=P|F\hb=P\Pi|\phi\hb.
\e{2311}
It follows that  $|C\hb$ is invariant under the general gauge transformations \rl{1231}, \ie
\be
&&|\phi\hb\quad\longrightarrow\quad|\phi\hb+\sum_{k=1}^sd_k^{\dag}|\varepsilon_k\hb.
\e{232}
The generalized Weyl transformations \rl{222} may then be given  the form
 \be
 &&|\phi\hb\;\;\longrightarrow\;\;|\phi\hb+\sum_{q,r=1}^sT^{\dag}_{qr}|\eta_{qr}\hb,
 \e{224}
where $|\eta_{qr}\hb$ is defined through
 \be
 &&|\zeta_{qr}\hb=\Pi|\eta_{qr}\hb
 \e{225}
 from \rl{231}. The index structure of $|\eta_{qr}\hb$ is determined by
 \be
 &&I^{(1)}_{qr}|{\eta}_{tu}\hb=\del_{qt}|{\eta}_{ur}\hb -\del_{qu}  |{\eta}_{tr}\hb
 \e{2321}
 from \rl{223}.
 
In $d=4$ \rl{224} is equivalent to  the generalized Weyl transformations given by Fradkin and Tseytlin in \cite{Fradkin:1985am}. There is no transformation of the gauge fields $|\phi\hb$ that yields  the dual Weyl transformations \rl{2302}.
 
\subsection{The Weyl states in terms of dual gauge fields}
Instead of expressing $|F\hb$ in terms of gauge fields $|\phi\hb$ one may express it in terms of the dual gauge fields $|\tilde{\phi}\hb$ according to (ignoring  the sign factor in \rl{1277})
\be
&&|F\hb=\Pi^{\dag}|\tilde{\phi}\hb,\quad \Pi^{\dag}=d_s\cdots d_1.
\e{233}
The properties of $|\tilde{\phi}\hb$ is given in subsection 3.2. $|C\hb$ is then invariant under the gauge transformations 
\be
&&|\tilde{\phi}\hb\quad\longrightarrow\quad|\tilde{\phi}\hb+\sum_{k=1}^sd_k|\tilde{\varepsilon}_k\hb.
\e{234}
and the dual generalized Weyl transformations follow then from \rl{2302}
 \be
 &&|\tilde{\phi}\hb\;\;\longrightarrow\;\;|\tilde{\phi}\hb+\sum_{q,r=1}^sT_{qr}|\tilde{\eta}_{qr}\hb,
 \e{235}
where $|\tilde{\eta}_{qr}\hb$ is defined through
 \be
 &&|\tilde{\zeta}_{qr}\hb=\Pi^{\dag}|\tilde{\eta}_{qr}\hb.
 \e{236}
The index structure of $|\tilde{\eta}_{qr}\hb$ is determined by
\be
&&I^{(-1)}_{qr}|\tilde{\eta}_{tu}\hb=\del_{rt}|\tilde{\eta}_{qu}\hb -\del_{ru}  |\tilde{\eta}_{qt}\hb
\e{2361}
from \rl{2301}. There is no transformation of the dual gauge fields $|\tilde{\phi}\hb$ that yields  the Weyl transformations \rl{222}.

 \setcounter{equation}{0}
\section{Weyl states for  half-integer spins}
\subsection{The projection operator $\cP$ for the Weyl states $|\cC\hb$}
For half-integer spins  I start with a spinor field $\Psi$ with the index structure according to condition (6) in \rl{113}:
\be
&&I_{qr}|\Psi\hb=0,\quad r,q=1,\ldots,[s].
\e{400}
The corresponding Weyl state $|\cC\hb$ is defined in \rl{202}. The 
 precise form of the projection operator $\cP$ in \rl{203}  is  up to third order ($\cP=1$ for $s=1/2$)
\be
&&\cP=1+\beta_1\sum_{r=1}^n\tau^{\dag}_r\tau_r+\beta_{21}\sum_{q,r=1}^n\tau^{\dag}_q\tau^{\dag}_r\tau_r\tau_q+\beta_{22}\sum_{q,r=1}^n\tau^{\dag}_q\tau^{\dag}_r\tau_q\tau_r+\nn\\&&+\beta_{31}\sum_{q,r,t=1}^n\tau^{\dag}_q\tau^{\dag}_r\tau^{\dag}_t\tau_t\tau_r\tau_q+\beta_{32}\sum_{q,r,t=1}^n\tau^{\dag}_q\tau^{\dag}_r\tau^{\dag}_t\tau_r\tau_t\tau_q+\nn\\&&+\beta_{33}\sum_{q,r,t=1}^n\tau^{\dag}_q\tau^{\dag}_r\tau^{\dag}_t\tau_r\tau_q\tau_t+
\beta_{34}\sum_{q,r,t=1}^n\tau^{\dag}_q\tau^{\dag}_r\tau^{\dag}_t\tau_q\tau_r\tau_t+\nn\\&&+
\beta_{35}\sum_{q,r,t=1}^n\tau^{\dag}_q\tau^{\dag}_r\tau^{\dag}_t(\tau_t\tau_q+\tau_q\tau_t)\tau_r+
O((\tau^{\dag})^4\tau^4),
\e{401}
where $n\equiv[s]$. The conditions for this expansion is 1) hermiticity $\cP=\cP^{\dag}$, 2) all indices are summed over. 3)  indices are always twice repeated: one index on a $\tau$ and one on a $\tau^{\dag}$. (The special form of the $\beta_{35}$-term is determined by hermiticity.) The maximal order $M$ is determined by the property
\be
&&(\tau_a)^{d/2+1}|\Psi\hb=0
\e{402}
for any fixed  index value $a$.

 \beq
 \begin{tabular}{|p{114mm}|}
 \hline
  {\bf Proposal 4}: The real parameters $\beta_{ki}$ in \rl{401} are determined by the conditions
 \be
 &&\tau_{q}|\cC\hb=0,\quad q=1,\ldots,[s].
 \e{403}\\
  \hline
 \end{tabular}
 \eq

 By means of the algebra \rl{112} I find to lowest orders ($n\equiv[s]$)
 \be
 &&\tau_{r}|\cC\hb=\Big(1+\beta_1 n\Big)\tau_{r}|\Psi\hb+\nn\\&&+\Big(\beta_1+\beta_{21} n+\beta_{22}(n+1)\Big)\sum_{r,t=1}^n\tau^{\dag}_{t}\tau_{r}\tau_{t}|\Psi\hb+\nn\\&&+
\Big(\beta_{21}(n+1)+\beta_{22}n\Big)\sum_{r,t=1}^n\tau^{\dag}_{t}\tau_{t}\tau_{r}|\Psi\hb+\nn\\&&+O\Big((\tau^{\dag})^2\tau^3\Big)|\Psi\hb.
 \e{404}
 Hence, condition \rl{403} yields ($n\equiv[s]$)
 \be
 &&\beta_1=-{1\over n},\nn\\
 &&\beta_{21}=-{1\over2n+1},\nn\\
 &&\beta_{22}={n+1\over n(2n+1)}.
  \e{405}
  This solution is valid for all even dimensions and all $[s]\geq1$.
  
  Notice that conditions \rl{403} imply
  \be
  &&T_{qr}|\cC\hb=0,\quad  q,r=1,\ldots,[s],
  \e{406}
  from the algebra \rl{112}.
  
  \subsection{The generalized Weyl invariance}
   \beq
 \begin{tabular}{|p{114mm}|}
 \hline
  {\bf Proposal 5}: The Weyl state satisfying $|\cC\hb=\cP|\Psi\hb$ is also invariant under the transformation ($n\equiv [s]$)
    \be
  &&|\Psi\hb\;\;\longrightarrow\;\;|\Psi\hb+\sum_{r=1}^n\tau^{\dag}_{r}|\zeta_{r}\hb,
  \e{407}
  where the  state $|\zeta_{r}\hb$ has the index structure
  \be
  &&I_{qr}|\zeta_{t}\hb=-\del_{qt}|\zeta_{r}\hb.
  \e{408}
  Vice versa if $|\cC\hb$ is invariant under \rl{407} then $|\cC\hb$ satisfies \rl{403}.\\
 \hline
 \end{tabular}
 \eq
Up to second order I find ($n\equiv [s]$)
\be
&&\sum_{r=1}^n\cP\tau^{\dag}_r|\zeta_r\hb=(1+\beta_1 n)\sum_{r=1}^n\tau_r^{\dag}|\zeta_r\hb+\nn\\&&+\big(\beta_1+\beta_{21} n +\beta_{22}(n+1)\big)\sum_{r,t=1}^n\tau^{\dag}_t\tau^{\dag}_r\tau_t|\zeta_r\hb+\nn\\&&+
\big(\beta_{21} (n+1) +\beta_{22} n\big)\sum_{r,t=1}^n\tau^{\dag}_r\tau^{\dag}_t\tau_t|\zeta_r\hb+
O\Big((\tau^{\dag})^3\tau^2|\zeta\hb\Big),
\e{409}
which vanishes for the values \rl{405}.

 \subsection{The index structure}
  \beq
 \begin{tabular}{|p{114mm}|}
 \hline
  {\bf Theorem 2}: The projection operator $\cP$ satisfies the property
  \be
  &&[I_{qr}, \cP]=0,\quad  q,r=1,\ldots,[s].
  \e{412}\\
 \hline
 \end{tabular}
 \eq
 
 Proof:  Consider the transformation of one index pair (t)
 \be
 &&[I_{qr}, \sum_t\cdots \tau^{\dag}_{t}\cdots \tau_{t}\cdots]=\nn\\
 &&\sum_t\Big\{\cdots (\del_{st}\tau^{\dag}_{r})\cdots \tau_{t}\cdots)+\cdots \tau^{\dag}_{t}\cdots (-\del_{rt}\tau_{s})\cdots)\Big\}+\cdots\nn\\
 &&=0+\cdots,
 \e{413}
 where the dots represents the terms coming from the transformation of all the other index pairs apart from $t$. Since one pair yields zero all pairs yield zero $\bullet$
 
 It follows that
 \be
 &&I_{qr}|\cC\hb=0,
 \e{414}
 since $I_{qr}|\Psi\hb=0$. Thus, the Weyl tensor $\cC$ has the same index structure as $\Psi$ for arbitrary even $d$. 
 
\subsection{$\cP$ as a projection operator}
From the form of $\cP$ in \rl{401} and the property \rl{403} it follows that 
\be
&&\cP|\cC\hb=|\cC\hb.
\e{415}
$P$ should therefore satisfy the projection property
\be
&&\cP^2=\cP+\sum_{r,q=1}^{[s]}\Big(\quad\Big)_{rq}I_{rq}.
\e{416}
Since this relation is nonlinear in the $\beta$-parameters, it is not useful to determine the $\beta$'s by the ansatz \rl{401}. (One solution of \rl{416} is of course also $\cP=1$.)

 \subsection{The dual projector $\tilde{\cP}$}
 The dual to $\cP$, $\tilde{\cP}$,  is according to appendix B equal to $\cP$ with the replacement $\tau \leftrightarrow -\tau^{\dag}$, which implies that $\tilde{\cP}$ has the structure (cf \rl{203})
 \be
&&\tilde{\cP}\equiv 1+\sum_{n=1}^N\sum_i\beta_{ni}(\tau)^n_{(i)}(\tau^{\dag})^n_{(i)}.
\e{417}
  \beq
 \begin{tabular}{|p{114mm}|}
 \hline
  {\bf Proposal 6}: The Weyl state $|\cC\hb$ in \rl{202} may also be written as
      \be
  &&|\cC\hb=\tilde{\cP}|\Psi\hb.
    \e{418}\\
 \hline
 \end{tabular}
 \eq
 This property requires that
 \be
 \sum[(\tau^{\dag})^n, \tau^n]=\sum_{q,r}\big(\cdots\big)_{qr}I_{qr}
 \e{419}
for each term in $\cP$ ($\tilde{\cP}$), which I have explicitly checked up to second order.

$\tilde{\cP}$ should also be a projector
\be
&&\tilde{\cP}|\cC\hb=|\cC\hb
\e{420}
satisfying
\be
&&\tilde{\cP}^2=\tilde{\cP}+\sum_{r,q=1}^{[s]}\Big(\quad\Big)_{rq}I_{rq}.
\e{4201}

A consequence of the relation \rl{418} is that the $\al$-parameters are also determined  by the dual conditions to \rl{403} and \rl{409}
 \be
 &&\tau^{\dag}_r|\cC\hb=0,\quad r=1,\ldots,[s],\nn\\
 &&\sum_{r=1}^{[s]}\cP\tau_r|\tilde{\zeta}_r\hb=0.
\e{4202}
The last property follows from the invariance of $|\cC\hb$ under the dual generalized Weyl transformations
\be
&&|\Psi\hb\quad\rightarrow\quad|\Psi\hb+\sum_{r=1}^{[s]}\tau_r|\tilde{\zeta}_r\hb,
\e{4203}
where
\be
&&I_{qr}|\tilde{\zeta}_t\hb=\del_{rt}|\tilde{\zeta}_q\hb.
\e{4204}
 
  From the relation \rl{418} it follows that the dual of $|\cC\hb$ satisfies the same properties as $|\cC\hb$. Thus, as on $|\Psi\hb$ one may impose (anti)self-duality conditions on $|\cC\hb$.
  
  \subsection{The Weyl states in terms of gauge fields}
  In order to express the Weyl states $|\cC\hb$ in terms of gauge fields I make use of the definition \rl{120} \ie
  \be
&&|\Psi\hb=\Pi |\rho\hb,\quad \Pi\equiv\Pi_{r=1}^{[s]} d^{\dag}_r\equiv d^{\dag}_1d^{\dag}_2\cdots d^{\dag}_{[s]}.
\e{4205}
From \rl{202} it follows then that
\be
&&|\cC\hb\equiv \cP|\Psi\hb=\cP\Pi|\rho\hb
\e{4206}
is invariant under the gauge transformations \rl{1231}
\be
&&|\rho\hb\;\;\longrightarrow\;\;|\rho\hb+\sum_{r=1}^{[s]}d^{\dag}_{r}|\xi_{r}\hb.
\e{42061}
Furthermore, from \rl{4206} and \rl{407} it follows that 
 $|\cC\hb$ is invariant under the generalized Weyl transformations
 \be
 &&|\rho\hb\;\;\longrightarrow\;\;|\rho\hb+\sum_{r=1}^{[s]}\tau^{\dag}_{r}|\eta_{r}\hb,
 \e{410}
where $|\eta_{r}\hb$ satisfies
 \be
 &&|\zeta_r\hb=\Pi|\eta_r\hb,\quad I^{(1)}_{qr}|\eta_r\hb=-\del_{qt}|\eta_r\hb,\quad r=1,\ldots,n.
 \e{411}
 Thus, \rl{410} implies the generalized Weyl transformations \rl{407}.
In $d=4$ \rl{410} is equivalent to  the generalized Weyl transformations given by Fradkin and Tseytlin in \cite{Fradkin:1985am}. There is no transformation of $|\rho\hb$ that yields dual generalized Weyl transformations \rl{4203}
 
 \subsection{The Weyl states in terms of dual gauge fields}
 If I express $|\Psi\hb$ in terms of dual gauge fields $|\tilde{\rho}\hb$ as in \rl{1271} using \rl{1274} \ie
\be
&&|\Psi\hb=\Pi^{\dag} |\cro\hb,\quad \Pi^{\dag}=\Pi_{r=1}^{[s]} d_r\equiv d_{[s]}d_{[s]-1}\cdots d_2d_1,
\e{4207}
then
\be
&&|\cC\hb=\cP|\Psi\hb=\cP\Pi^{\dag}|\trho\hb
\e{4208}
is invariant under the generalized gauge transformations \rl{1231}
\be
&&|\trho\hb\quad\rightarrow\quad|\trho\hb+\sum_{r=1}^{[s]}d_r|\tve_r\hb,
\e{4209}
and the dual generalized Weyl transformations
\be
&&|\trho\hb\quad\rightarrow\quad|\trho\hb+\sum_{r=1}^{[s]}\tau_r|\tilde{\eta}_r\hb,
\e{4210}
where
\be
&&|\tilde{\zeta}_r\hb=\Pi^{\dag}|\tilde{\eta}_r\hb,
\e{4211}
which makes \rl{4210} equivalent to \rl{4203}. Relations \rl{4210} and \rl{4211} requires the index structure
\be
&&I_{rq}^{(-1)}|\tilde{\eta}_t\hb=\del_{tq}|\tilde{\eta}_r\hb.
\e{4212}
 There is no transformation of $|\tilde{\rho}\hb$ that yields generalized Weyl transformations \rl{407}.

 \setcounter{equation}{0}
\section{Gauge invariant actions and equations for the wave functions}
The theory considered so far have no equations of motion. The equations of the exact theory, \ie (1)-(3),(7) in \rl{113},\rl{114}, were removed from the start. Instead the appropriate equations are here introduced as Lagrangian equations derivable from  actions for the wave functions. These actions are simply 
\be
&&S_1\equiv \vb C|C\hb,\quad S_2\equiv\vb \cC|\cp|\cC\hb.
\e{421}
(Insertion of $\cp$ in $S_2$ is dictated by conformal invariance.)
These actions are gauge invariant under \rl{1231} when written in terms of the gauge fields $\phi$ and $\rho$  according to \rl{120}. In addition, they are invariant under the generalized Weyl transformations \rl{224}  and \rl{410}. The expressions in \rl{421} yield
\be
&&S_1\equiv \vb C|C\hb=\vb F|P^{\dag}|C\hb=\vb F|P|C\hb=\vb F|C\hb=\vb\phi|\Pi^{\dag}|C\hb,\nn\\
&&S_2\equiv\vb \cC|\cp|\cC\hb=\vb \Psi|\cP^{\dag}\cp|\cC\hb=\vb \Psi|\cP\cp|\cC\hb=(1+[s])\vb \Psi|\cp|\cC\hb=\nn\\&&\quad\;=(1+[s])\vb\rho|\Pi^{\dag}\cp|\cC\hb,
\e{422}
where I have used the properties of the projection operators $P$ and $\cP$ and the definitions \rl{120} of the gauge fields $\phi$ and $\rho$. Thus, the equations of motion are (see \rl{1272})
\be
&&\Pi^{\dag}|C\hb= d_1d_2\cdots d_s|C\hb=0,\nn\\
&&\Pi^{\dag}\cp|\cC\hb=\cp d_1d_2\cdots d_{[s]}|\cC\hb=0,
\e{423}
which means that the order of the equations for the gauge fields $\phi$ and $\rho$ are $2s$. (These equations differ from those of the exact theory for $s>1$.) The above  results are valid irrespective of the reality of the basic fields $\phi$ and $\rho$. For real fields the action $S_1$ in \rl{421} and the corresponding equations of motion in \rl{423} agree with the results for integer spins $s$ in \cite{Marnelius:2008er}. One may of course also express the Weyl states in terms of dual gauge fields in which case the equations are \rl{423} with $d_k$ replaced by $d^{\dag}_k$. The actions \rl{421} when written in a standard form are (I drop a factor $\sqrt{2}$ in $S_2$ from \rl{1121})
\be
&&S_1=\int dx\; C^2(x),\nn\\
&&S_2=i\int dx\; \bar{\cC}(x)\cdif\cC(x),
\e{424}
where $C(x)$ and $\cC(x)$ are the corresponding Weyl tensors to $|C\hb$ and $|\cC\hb$ respectively.
In $d=4$ $S_1$ and $S_2$ seem to be in agreement with the actions proposed by Fradkin and Linetski in \cite{Fradkin:1989md,Fradkin:1990ps}. Notice, however, that the index symmetry of $C$ and $\cC$ are different here as compared with \cite{Fradkin:1989md,Fradkin:1990ps}, but they are at the same time equivalent  as explained in section 6 in \cite{Marnelius:2008er}.

 \setcounter{equation}{0}
\section{Final remarks on the higher order theories}
The compact treatment of the higher order theory given here elucidate and further extend the results of \cite{Marnelius:2008er}. Firstly, I have treated half-integer spins here which was lacking in \cite{Marnelius:2008er}. Secondly, the treatment here is valid in any even dimension  $d$ and not just $d=4$ which was the main part of   \cite{Marnelius:2008er}. Thirdly, the duality properties given here are also new. Notice that
\be
&&|C\hb=P|F\hb,\quad|\cC\hb=\cP|\Psi\hb,
\e{501}
implies
\be
&&|\tilde{C}\hb=\tilde{P}|\tilde{F}\hb,\quad |\tilde{\cC}\hb=\tilde{\cP}|\tilde{\Psi}\hb,
\e{502}
according to appendix B. From \rl{227} and \rl{418} it follows then that $|\tilde{C}\hb$ and $ |\tilde{\cC}\hb$ satisfy the same properties as $ |C\hb$ and $ |\cC\hb$ since $|\tilde{F}\hb/|\tilde{\Psi}\hb$ satisfy the same properties a $ |F\hb/|\Psi\hb$.   Thus, also the Weyl tensors may be chosen to be (anti)self-dual like $F$ and $\Psi$. Fourthly, the index structure of the Weyl tensors which was only conjectured for $d=4$ in section 6 in \cite{Marnelius:2008er},  are proved here for any even dimension $d$ and for both integer and half-integer spins in subsections 9.3 and 10.3. This proves the proposal 4 in \cite{Marnelius:2008er} in general. This means that the higher order theories given here are in $d=4$ in agreement with the general framework proposed by Fradkin,Linetsky and Tseytlin in \cite{Fradkin:1985am,Fradkin:1989md,Fradkin:1990ps} although the Weyl tensors here are in a different but equivalent representation as explained in appendix A in \cite{Marnelius:2008er}.

\setcounter{equation}{0}
\section{Comparison between the exact and the higher order theories}
The exact theory is  a second order theory for the gauge fields in the integer spin case and a first order one in the half-integer case if one ignores the fact that compensator fields involve higher orders. The limiting (Fang)-Fronsdal Lagrangians are strictly of (first) second order.   The fact that the compensator fields satisfy higher order equations (fourth and third orders) is probably no problem in itself.  However, there is perhaps another problem with the introduction of compensator fields and Lagrange multipliers: What is the very meaning of their presence  in the actions? At best they provide for an extended gauge symmetry. However, they may also imply the presence of second class constraints. If this is the case they must be solvable in a Lorentz covariant way. If not the Lagrangians cannot be used in a standard covariant quantization. It is therefore urgent to perform a constraint analysis of all Lagrangians for free higher spin fields  to determine whether or not there are second class constraints and/or additional gauge invariances. Maybe not all forms yield the same result. Apart from \cite{Francia:2005bu,Campoleoni:2008jq,Campoleoni:2009gs} and the form here there has recently appeared a different form in \cite{Sorokin:2008tf}. A good free Lagrangian is prerequisite to the much more difficult task to find interactions.

The Lagrangians for the higher order theories have no auxiliary variables and cannot involve any second class constraints. They are extended gauge theories containing apart from the full gauge invariance also generalized Weyl invariance. However, they are of higher order. Even though higher order gauge theories may be quantized in a formally consistent way they do in general involve indefinite metric states in the physical spectrum. It is urgent to clarify this point. Furthermore, it remains to determine what they actually describe since only the exact theory describes precisely massless particles with definite spins (in $d=4$). On the other hand, the advantage of the higher order theory is that it allows for interactions within a Lagrangian scheme. At least there are definite signs of this:  the consistent cubic interactions in \cite{Fradkin:1989md,Fradkin:1990ps}, and the consistent external field coupling  in   \cite{Arvidsson:2006}.

 \begin{appendix}
\newpage
\section{Duality properties of the basic Fock space}
Consider one set of  odd $b$-operators as defined in \rl{104}. They satisfy the anticommutation relations
   \be
   &&[b^{\mu}, b^{\dag\nu}]_+=\eta^{\mu\nu},\nn\\
   &&[b^{\mu}, b^{\nu}]_+=[b^{\dag\mu}, b^{\dag\nu}]_+=0,
   \e{a1}
   according to \rl{105}. 
   For these oscillators it is possible to define {\em two} different vacuum states:
   \be
   &&b^{\mu}|0\hb=0,\nn\\
   &&b^{\dag\mu}\widetilde{|0\hb}=0.
   \e{a2}
   These vacua may be related as follows
   \be
   &&D\widetilde{|0\hb}=|0\hb,\nn\\
   &&D\equiv C{1\over d!}\varepsilon^{\mu_1\mu_2\cdots\mu_d}b_{\mu_1}b_{\mu_2}\cdots b_{\mu_d},
   \e{a3}
   where $C$ is a complex constant and
where $\varepsilon$ is the totally antisymmetric tensor satisfying $\varepsilon^{012\cdots d-1}=1$. $D$ is an even operator for even dimensions $d$. In the following $\widetilde{|0\hb}$ will be referred to as the dual vacuum state to $|0\hb$. A straight forward calculation yields
\be
&&DD^{\dag}|0\hb=CC^*\Big(\Pi_{k=0}^{d-1}\eta_{kk}\Big)|0\hb=-CC^*|0\hb,\nn\\
&&D^{\dag}D\widetilde{|0\hb}=-CC^*\widetilde{|0\hb},
\e{a4}
if a Minkowski metric $\eta_{\mu\nu}$ is used. (A Euclidean metric yields  a plus sign.) The second line and \rl{a3} yield then
\be
&&\vb 0|0\hb=-CC^*\widetilde{\vb 0}|\widetilde{0\hb}.
\e{a5}
Thus, the vacuum and its dual have opposite norms.
It is natural to require the dual of the dual vacuum to be the original vacuum, \ie
\be
&&\widetilde{\widetilde{|0\hb}}=|0\hb.
\e{a6}
The relation \rl{a5} requires then
\be
&&CC^*=1,\quad\Leftrightarrow\quad\vb 0|0\hb=-\widetilde{\vb 0}|\widetilde{0\hb}.
\e{a7}
From \rl{a3} it is natural to define a dual operator to $D$ by
\be
&&\tilde{D}|0\hb=\widetilde{|0\hb}.
\e{a8}
This together with \rl{a3} yields
\be
&&D\tilde{D}|0\hb=|0\hb,\quad\tilde{D}D\widetilde{|0\hb}=\widetilde{|0\hb}.
\e{a9}
From \rl{a4} one finds then (using \rl{a7})
\be
&&\tilde{D}=-D^{\dag}.
\e{a10}

The Fock states are defined by
\be
&&|0\hb_{\mu_1\cdots\mu_k}\equiv {1\over\sqrt{k!}}b_{\mu_1}^{\dag}b_{\mu_2}^{\dag}\cdots b_{\mu_k}^{\dag}|0\hb.
\e{a11}
If
\be
&&_{\nu_1\cdots\nu_k}\vb 0|\equiv\Big(|0\hb_{\nu_1\cdots\nu_k}\Big)^{\dag},\quad\vb0|0\hb=1,
\e{a12}
then
\be
&&_{\nu_1\cdots\nu_k}\vb 0|0\hb_{\mu_1\cdots\mu_k}={1\over k!}\sum_{antisym \;\mu's}\eta_{\mu_1\nu_1}\cdots\eta_{\mu_k\nu_k}=\nn\\&&=-{(d-k)!\over k!}\varepsilon_{\nu_1\cdots\nu_d}\varepsilon_{\mu_1\cdots\mu_d}\eta^{\nu_{k+1}\mu_{k+1}}\cdots\eta^{\nu_d\mu_d}.
\e{a13}
Thus, if $F$ is a tensor operator which is independent of the $b$-operators then
\be
&&_{\nu_1\cdots\nu_k}\vb 0|F\hb=F_{\nu_1\cdots\nu_k},\nn\\
&&|F\hb\equiv F^{\mu_1\cdots\mu_k}|0\hb_{\mu_1\cdots\mu_k}.
\e{a14}
There is also a dual Fock space expressed in terms of the dual vacuum state $\widetilde{|0\hb}$.
I find the basic states to be
\be
&&\widetilde{|0\hb}_{\mu_1\cdots\mu_k}\equiv {1\over\sqrt{k!}}b_{\mu_1}b_{\mu_2}\cdots b_{\mu_k}\widetilde{|0\hb}=
-{1\over C}f_{d,k}\;\varepsilon_{\nu_1\cdots\nu_{d-k}\mu_1\cdots\mu_k}|0\hb^{\nu_{d-k}\cdots\nu_1},\nn\\
&&|0\hb_{\mu_1\cdots\mu_k}\equiv {1\over\sqrt{k!}}b^{\dag}_{\mu_1}b^{\dag}_{\mu_2}\cdots b^{\dag}_{\mu_k}|0\hb=
Cf_{d,k}\;\varepsilon_{\mu_k\cdots\mu_1\nu_{k+1}\cdots\nu_{d}}\widetilde{|0\hb}^{\nu_{k+1}\cdots\nu_{d}},\nn\\
\e{a15}
where
\be
&&f_{d,k}\equiv{1\over\sqrt{k!(d-k)!}}.
\e{a16}
This implies  the following relations for  a general state
\be
&&|F\hb\equiv F^{\mu_1\cdots\mu_k}|0\hb_{\mu_1\cdots\mu_k}=Cf_{d,k}F^{\mu_1\cdots\mu_k}\varepsilon_{\mu_k\cdots\mu_1\nu_{k+1}\cdots\nu_d}\widetilde{|0\hb}^{\nu_{k+1}\cdots\nu_d},\nn\\
&&\widetilde{|F\hb}\equiv F^{\mu_1\cdots\mu_k}\widetilde{|0\hb}_{\mu_1\cdots\mu_k}=-{1\over C}f_{d,k}F^{\mu_1\cdots\mu_k}\varepsilon_{\nu_{1}\cdots\nu_{d-k}\mu_1\cdots\mu_k}|0\hb^{\nu_{d-k}\cdots\nu_1}=|\tilde{F}\hb.\nn\\
\e{a17}
It would be natural to define dual fields $\tilde{F}$ by
\be
&&|F\hb\equiv\widetilde{|\tilde{F}\hb},\quad{\rm or}\quad\widetilde{|F\hb}\equiv|\tilde{F}\hb,
\e{a18}
which are explicitly given by
\be
&&F^{\mu_1\cdots\mu_k}|0\hb_{\mu_1\cdots\mu_k}\equiv    \tilde{F}^{\mu_1\cdots\mu_k}  \widetilde{|0\hb}_{\mu_1\cdots\mu_k}\nn\\&& F^{\mu_1\cdots\mu_k}\widetilde{|0\hb}_{\mu_1\cdots\mu_k}\equiv\tilde{F}^{\mu_1\cdots\mu_k}|0\hb_{\mu_1\cdots\mu_k}.
\e{a181}
The first definition implies
\be
&&\tilde{F}_{\nu_1\cdots\nu_{d-k}}=Cf_{d,k}F^{\mu_1\cdots\mu_k}\varepsilon_{\mu_k\cdots\mu_1\nu_1\cdots\nu_{d-k}},
\e{a19}
while the second definition yields
\be
&&\tilde{F}_{\nu_1\cdots\nu_{d-k}}=-{1\over C}f_{d,k}F^{\mu_1\cdots\mu_k}\varepsilon_{\nu_{d-k}\cdots\nu_{1}\mu_1\cdots\mu_k},
\e{a20}
from \rl{a17}. Notice, however, that the second definition coincide with the first if $F$ is replaced by $\tilde{F}$ according to \rl{a18}. Both expressions \rl{a19} and \rl{a20} yield the same dual field $\tilde{F}$ if the constant $C$ is chosen to be (two choices)
\be
&&C=\pm i(-1)^{{d(d-1)\over4}},
\e{a21}
which is consistent with the condition \rl{a7}. In the above definition $\tilde{F}$ is not required to be real when $F$ is real. In fact, $F$ and $\tilde{F}$ may be both real only in dimensions $d=4n+2$. (For $d=4n$ reality requires $\tilde{\tilde{F}}=-F$.)

The index operator \rl{1221}  is here given by
\be
&&I^{(n)}\equiv\half\Big(b^{\dag}\cdot b-b\cdot b^{\dag}\Big)+n=b^{\dag}\cdot b+(n-d/2).
\e{a22}
The vacuum states \rl{a2} satisfy
\be
&&I^{(d/2)}|0\hb,\quad I^{(-d/2)}\widetilde{|0\hb}=0,
\e{a23}
which are consistent with \rl{a3} and \rl{a8} since
\be
&&[I^{(n)}, D]=-dD,\quad [I^{(n)}, \tilde{D}]=d\tilde{D}
\e{a24}
for any value of $n$. The base states \rl{a15} satisfy
\be
&&I^{(d/2-k)}|0\hb_{\mu_1\cdots\mu_k}=0,\quad I^{(k-d/2)}\widetilde{|0\hb}_{\mu_1\cdots\mu_k}=0.
\e{a25}
The above expressions make it natural to define the dual index operator to \rl{a22} by
\be
&&\tilde{I}^{(n)}\equiv -\half\Big(b^{\dag}\cdot b-b\cdot b^{\dag}\Big)+n=-I^{(-n)},
\e{a251}
which is consistent with the general definitions of dual operators in appendix B.

In the text a set of $[s]$ $b$-operators are used. In this case one has to define a total vacuum state by
\be
&&|0\hb\equiv\Pi_{k=1}^{[s]}|0\hb_k,\quad \widetilde{|0\hb}\equiv\Pi_{k=1}^{[s]}\widetilde{|0\hb}_k.
\e{a26}
(Even various mixtures are possible.) The above vacua are related by
\be
&&D\widetilde{|0\hb}=|0\hb,\quad D\equiv\Pi_{k=1}^{[s]}D_k,\quad \vb0|0\hb=(-1)^{[s]}\widetilde{\vb 0}\widetilde{|0\hb}.
\e{a27}
The general Fock space is then spanned by
\be
&&|0\hb^{\mu_1\cdots \mu_{n_1};\nu_1\cdots \nu_{n_2};\rho_1\cdots \rho_{n_3};\cdots;\la_1\cdots \la_{n_s}}
\equiv  {1\over\sqrt{n_1!n_2!\cdots n_s!}}\times\nn\\
&& b_1^{\mu_1\dagger}\cdots b_1^{\mu_{n_1}\dagger}b_2^{\nu_1\dagger}\cdots b_2^{\nu_{n_2}\dagger}b_3^{\rho_1\dagger}\cdots b_3^{\rho_{n_3}\dagger}\cdots\cdots b_s^{\la_{1}\dagger}\cdots b_s^{\la_{n_s}\dagger} |0\hb,\nn\\
&&\widetilde{|0\hb}^{\mu_1\cdots \mu_{n_1};\nu_1\cdots \nu_{n_2};\rho_1\cdots \rho_{n_3};\cdots;\la_1\cdots \la_{n_s}}
\equiv  {1\over\sqrt{n_1!n_2!\cdots n_s!}}\times\nn\\
&& b_1^{\mu_1}\cdots b_1^{\mu_{n_1}}b_2^{\nu_1}\cdots b_2^{\nu_{n_2}}b_3^{\rho_1}\cdots b_3^{\rho_{n_3}}\cdots\cdots b_s^{\la_{1}}\cdots b_s^{\la_{n_s}} \widetilde{|0\hb}.
\e{a28}

\setcounter{equation}{0}
\section{General duality properties}
The dual operators to the basic operators in the quantum supersymmetric particle model considered in the text are defined as follows:
\be
&&\widetilde{b_{k}^{\mu}}={b}_{k}^{\dag\mu},\quad\widetilde{b_{k}^{\dag\mu}}={b}_{k}^{\mu},\quad\tilde{x}=x,\quad\tilde{p}=p,\quad\tilde{\psi}=\psi.
\e{b1}
For an arbitrary operator $G$ the dual operator $\tilde{G}$ is required to satisfy
\be
&&\tilde{\tilde{G}}=G,
\e{b2}
which is satisfied by the operators in \rl{b1}. As a general rule I propose that  dual operators are simply defined by interchanging $b$ and $b^{\dag}$, \ie
\be
&&{b}_{k}^{\mu}\quad\longleftrightarrow\quad{b}_{k}^{\dag\mu}.
\e{b3}
When this rule is applied to the basic index operator\rl{111} I find 
\be
&&\tI_{rq}=\half\Big(b_r\cdot b^{\dag}_q-b^{\dag}_q\cdot b_r\Big)=-I_{qr}.
\e{b4}
For the more general index operator in \rl{1221} I have then
\be
&&\tI^{(n)}_{rq}=-I_{qr}+n\del_{rq}=-I_{qr}^{(-n)},
\e{b5}
which agrees with \rl{a251} for $r=q=1$. Notice that $\tI$ is not even linearly related to the hermitian conjugate of $I$. Dual operators can therefore not be defined by simply hermitian conjugation. Notice also that the operator $\tilde{D}$ defined in appendix A is also obtained from $D$ by the rule \rl{b3}. In the main text apart from \rl{b1} and \rl{b5} the following dual operators are used
\be
&&\tilde{d}_r=d^{\dag}_r,\nn\\
&&\tilde{T}_{rq}=-T^{\dag}_{rq},\nn\\
&&\tilde{\tau}_r=-\tau^{\dag}_r,
\e{b6}
all obtained from the rule \rl{b3}.

For the states duality simply means the interchange of the two vacua in \rl{a2} apart from the above rule. For general states it is also natural to require
\be
&&\widetilde{\widetilde{|A\hb}}=|A\hb.
\e{b7}
For general states like the ans{\"a}tze \rl{106} and \rl{108} I require
\be
&&\widetilde{|F\hb}=|\tilde{F}\hb,\quad \widetilde{|\tilde{F}\hb}=|F\hb,\nn\\
&&\widetilde{|\Psi\hb}=|\tilde{\Psi}\hb,\quad \widetilde{|\tilde{\Psi}\hb}=|\Psi\hb,
\e{b8}
which is consistent with what was found in appendix A. Notice that relations like 
\be
&&G_1|F\hb=0,\quad G_2|\Psi\hb=0,
\e{b9}
are equivalent to the relations
\be
&&\tilde{G}_1|\tilde{F}\hb=0,\quad \tilde{G}_2|\tilde{\Psi}\hb=0,
\e{b10}
from \rl{b7} and \rl{b8}. This is used in the main text to relate equations for the dual fields to the equations for the fields.

\setcounter{equation}{0}
\section{The gauge invariant Lagrangian equations in the exact case with general compensator fields}
The action $S_1$ in \rl{174} yields the first equations in \rl{190} and \rl{1903} together with
\be
&&\half\Big(1-{1\over4}T_{ut}^{\dag}T_{ut}\Big)|E_{\phi}\hb+\half\Big(p^2-d_r^{\dag}d_r-\half T^{\dag}_{rq}d_qd_r\Big)\big(1-{1\over4}T_{ut}^{\dag}T_{ut}\Big)|{\phi}\hb+\nn\\&&+{1\over8}\Big(p^2-d_r^{\dag}d_r-\half T^{\dag}_{rq}d_qd_r\Big)d_v^{\dag}T^{\dag}_{ut}|\al_{utv}\hb+{1\over24}(T^{\dag}T^{\dag})_{uqtr}|\beta_{uqtr}\hb=0\nn\\
\e{d1}
from the variation of $\phi$, and
\be
&&{1\over8}\Big(T_{qt}d_r+T_{tr}d_q+T_{rq}d_t\Big)|E_{\phi}\hb-{1\over6}d^{\dag}_k|\beta_{kqtr}\hb-\nn\\&&-{1\over24}T^{\dag}_{ku}\Big\{2d_u|\beta_{kqtr}\hb+d_r|\beta_{kuqt}\hb+d_q|\beta_{kutr}\hb+d_t|\beta_{kurq}\hb\Big\}-\nn\\&&-{1\over12}d_qd_td_r\Big\{\Big(1-{1\over4}T_{uk}^{\dag}T_{uk}\Big)|\phi\hb+d_v^{\dag}T^{\dag}_{uk}|\al_{ukv}\hb\Big\}=0
\e{d2}
from the variation of $\la_{qtr}$.

The action $S_2$ in \rl{174} yields the second equations in \rl{190} and \rl{1903} together with
\be
&&\half\Big(1-\tau_k^{\dag}\tau_k+{1\over3}\tau^{\dag}_k\tau^{\dag}_q(T_{qk}+\tau_k\tau_q)\Big)|E_{\rho}\hb+\nn\\&&+\half(\cp-\tau^{\dag}_rd_r)\Big\{\Big(
1-\tau_k^{\dag}\tau_k+{1\over3}(T^{\dag}_{qk}+\tau_q^{\dag}\tau_k^{\dag})\tau_q\tau_k\Big)|\rho\hb+\nn\\&&+\Big(\tau^{\dag}_kd_r^{\dag}+{1\over3}(\tau^{\dag}_r\tau_k^{\dag}+\tau_k^{\dag}\tau_r^{\dag})\cp\Big)|\al_{kr}\hb-{1\over3}d_r^{\dag}(T^{\dag}_{qk}+\tau_q^{\dag}\tau_k^{\dag})\tau_q|\al_{kr}\hb\Big\}-\nn\\&&-{1\over12}(\tau^{\dag}T^{\dag})_{tkr}|\beta_{tkr}\hb=0
\e{d3}
from the variation of $\rho$, and
\be
&&\Big(d_t\tau_r+{1\over3}\tau^{\dag}_q(T_{qt}+\tau_t\tau_q)d_r-(r\leftrightarrow t)\Big)|E_{\rho}\hb+\nn\\&&+d_td_r\Big(1-\tau^{\dag}_k\tau_k+{1\over3}(T^{\dag}_{qk}+\tau^{\dag}_q\tau^{\dag}_k)\tau_q\tau_k\Big)|\rho\hb+\nn\\&&+d_td_r\Big(\tau^{\dag}_kd^{\dag}_u +{1\over3}(\tau^{\dag}_u \tau_k^{\dag}+\tau_k^{\dag}\tau^{\dag}_u)\cp\Big)|\al_{ku}\hb-{1\over3}d_td_rd^{\dag}_u(T^{\dag}_{qk}+\tau_q^{\dag}\tau^{\dag}_k)\tau_q|\al_{ku}\hb+\nn\\&&+d_k^{\dag}|\beta_{rtk}\hb+\tau^{\dag}_k\cp|\beta_{rtk}\hb-\tau_k^{\dag}\tau_q^{\dag}d_q|\beta_{rtk}\hb+\half T^{\dag}_{kq}\Big(d_t|\beta_{rkq}\hb-d_r|\beta_{tkq}\hb\Big)=0\nn\\
\e{d4}
from the variation of $\la_{rt}$.

\end{appendix}

\bibliographystyle{utphysmod2}
\bibliography{biblio}
\end{document}